\documentclass[10pt,conference,letterpaper,nofonttune]{IEEEtran}
\usepackage{fancyhdr}

\makeatother

\usepackage[acronyms]{glossaries}
\newacronym{dl}{DL}{deep learning}
\newacronym{dnn}{DNN}{deep neural network}

\newacronym{cnn}{CNN}{convolutional neural network}
\newacronym{csi}{CSI}{Channel State Information}
\newacronym{cv}{CV}{computer vision}
\newacronym{har}{HAR}{human activity recognition}
\newacronym{lan}{LAN}{local-area network}
\newacronym{lstm}{LSTM}{long short-term memory}
\newacronym{mimo}{MIMO}{multiple-input multiple-output}
\newacronym{nic}{NIC}{network interface card}
\newacronym{ofdm}{OFDM}{orthogonal frequency-division multiplexing}
\newacronym{ofdma}{OFDMA}{orthogonal frequency-division multiple access}
\newacronym{siso}{SISO}{single-input single-output}
\newacronym{sta}{STA}{station}
\newacronym{wlan}{WLAN}{wireless local-area network}

\newacronym{bfi}{BFI}{beamforming feedback information}

\newacronym{mum}{MU-MIMO}{multi-user multi-input multi-output}

\newacronym{fsl}{FSL}{Few-Shot Learning}
\newacronym{phy}{PHY}{physical layer}

\newacronym{snr}{SNR}{signal-to-noise ratio}
\newacronym{ndp}{NDP}{null data packet}
\newacronym{svd}{SVD}{singular value decomposition}

\newacronym{frel}{FREL}{Feature Reusable Embedding  Learning}

\newacronym{fsel}{FSEL}{few-shot embedding learning}
\newacronym[plural=\gls{ltf}s,firstplural=long training fields (LTFs)]{ltf}{LTF}{long training field}

\newacronym{free}{\texttt{FReE}}{\textit{Feature Reusable Embedding Learning}}
\newacronym{relu}{ReLU}{Rectified Linear Unit}
\newacronym{knn}{K-NN}{K-Nearest Neighbor}
\newacronym{ap}{AP}{Access Point}
\usepackage{pgfplots}
\usepackage{tikz}
\usepackage{pgfplotstable}
\usepackage[utf8]{inputenc}
\pgfplotsset{compat=1.17}
\newlength\fheight
\newlength\fwidth
\usetikzlibrary{plotmarks,patterns,decorations.pathreplacing,backgrounds,calc,arrows,arrows.meta,spy,matrix}
\usepgfplotslibrary{patchplots,groupplots,colorbrewer}
\usepackage{tikzscale}

\usepackage{subcaption}
\usepackage{caption}

\usepackage{algorithm2e}
\newcommand{\FW}{\texttt{SiMWiSense}\xspace}
\newcommand{\FR}{\gls{frel}\xspace}

\newcommand{\ap}{\gls{ap}\xspace}

\usepackage{cite}
\usepackage{amsmath,amssymb,amsfonts}
\usepackage{algorithmic}
\usepackage{graphicx}
\usepackage{textcomp}
\usepackage{xcolor}
\usepackage[]{footmisc}

\def\BibTeX{{\rm B\kern-.05em{\sc i\kern-.025em b}\kern-.08em
    T\kern-.1667em\lower.7ex\hbox{E}\kern-.125emX}}

\begin{document}

\newcommand{\fr}[1]{\begin{color}{red}\textbf{[FR: #1]}\end{color}}

\title{\FW: Simultaneous Multi-Subject \\ Activity Classification Through Wi-Fi Signals}

\author{\IEEEauthorblockN{Khandaker Foysal Haque, Milin Zhang, and Francesco Restuccia}
\IEEEauthorblockA{ Institute for the Wireless Internet of Things, Northeastern University, United States\\
\{haque.k, zhang.mil, f.restuccia\}@northeastern.edu}\vspace{-1cm}
}

\maketitle

\begin{abstract}
Recent advances in Wi-Fi sensing have ushered in a plethora of pervasive applications in home surveillance, remote healthcare, road safety, and home entertainment, among others. Most of the existing works are limited to the activity classification of a single human subject at a given time. Conversely, a more realistic scenario is to achieve \textit{simultaneous, multi-subject} activity classification. The first key challenge in that context is that the number of classes grows exponentially with the number of subjects and activities. Moreover, it is known that Wi-Fi sensing systems struggle to adapt to new environments and subjects. To address both issues, we propose \FW, the first framework for simultaneous multi-subject activity classification based on Wi-Fi that generalizes to multiple environments and subjects. We address the scalability issue by using the \gls{csi} computed from the device positioned closest to the subject. We experimentally prove this intuition by confirming that the best accuracy is experienced when the \gls{csi} computed by the transceiver positioned closest to the subject is used for classification. To address the generalization issue, we develop a brand-new few-shot learning algorithm named \FR. Through an extensive data collection campaign in 3 different environments and 3 subjects performing 20 different activities simultaneously, we demonstrate that \FW achieves classification accuracy of up to 97\%, while \FR improves the accuracy by 85\% in comparison to a traditional Convolutional Neural Network (CNN) and up to 20\% when compared to the state-of-the-art \gls{fsel}, by using only 15 seconds of additional data for each class. For reproducibility purposes, we share our 1TB dataset and code repository\footnote{https://github.com/kfoysalhaque/SiMWiSense}\cite{SiMWiSense}.
\end{abstract}

\glsresetall

\section{Introduction}

Wi-Fi is one of the most pervasive wireless technologies worldwide -- it has been estimated that by 2025, the Wi-Fi economy  will reach a value of \$4.9T \cite{WiFiAlliance}. Beyond ubiquitous indoor connectivity, Wi-Fi also allows to develop highly-pervasive device-free sensing applications. The latter are based on the intuition that the received Wi-Fi signals -- in particular, the \gls{csi} computed to perform channel estimation and equalization -- are affected by changes in the physical environment caused by any entity in between the source and the receiver. Among other applications, Wi-Fi sensing can be used for fine-grained indoor localization \cite{8767421}, activity recognition \cite{9762352, 9439116}, and health monitoring \cite{zeng2020multisense}. For an excellent survey on the topic, we refer the reader to \cite{ma2019wifi}.

Most of the relevant existing work -- discussed in detail in Section \ref{sec:related_work} -- focuses on performing classification of a single subject at a given time \cite{abdelnasser2018ubiquitous,qian2017widar, meneghello2022sharp,wang2017phasebeat}. Even though achieving acceptable sensing performance, a significantly more relevant (and realistic) problem is performing \textit{simultaneous, multi-subject} Wi-Fi sensing. Moreover, it is well known that Wi-Fi sensing is highly-dependent of the considered subject and environment  \cite{ma2021location}. Although some attempts to address the issue have been made, they consider few activities -- less than 5  \cite{bahadori2022rewis} -- or do not consider multi-subject classification \cite{meneghello2022sharp}.

\begin{figure}[h]
	\centering
	\includegraphics[width=\columnwidth]{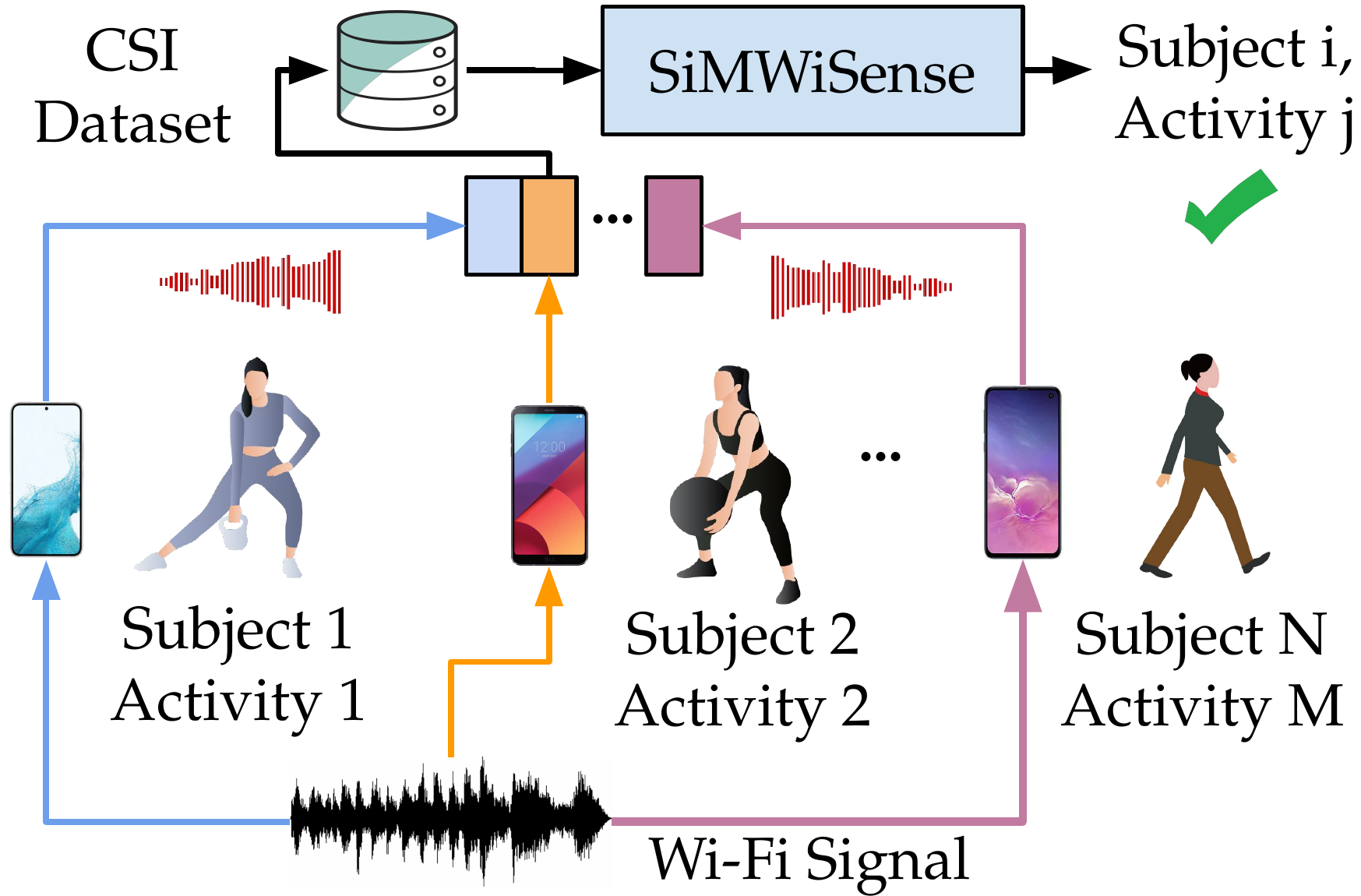}
	\caption{High-level overview of \FW.}
	\label{fig:simsense}
\end{figure}

In stark contrast to the existing works, we propose \FW, a completely novel approach for simultaneous multi-subject activity classification through Wi-Fi.  Figure \ref{fig:simsense} shows a high-level overview of our approach. Beyond the generalization issue, the key challenge addressed by \FW is that by defining as $n$ and $m$ respectively the number of subjects and activities, the number of classes to distinguish becomes $n^m$. For example, 3 subjects and 10 activities correspond to more than 59,000 classes.

To address this critical issue, we utilize multiple Wi-Fi devices as CSI collectors, where the \textit{closest} to a given subject will  classify the activities conducted by that subject. We experimentally prove in Section \ref{sec:prox_test} that the device closest to the subject will dominantly characterize the channel property between itself and the source of the Wi-Fi signal. Finding the closest device to a given subject falls under the Wi-Fi indoor localization and/or fingerprinting problem \cite{shokry2017tale,abbas2019wideep,wang2019joint}, which has been thoroughly investigated and thus considered out of our scope. Although assigning a device to a subject addresses the scalability issue -- the classifier output becomes $m$ sized -- the overall performance may significantly degrade with new untrained environments and subjects. Thus, we developed a novel \gls{fsl} architecture which can adapt to any new environment, change in environment or any new subject with up to 15 seconds of new data for each class.

\subsection*{\textbf{Summary of Novel Contributions}}
$\bullet$ We present \FW, the first framework for multi-subject simultaneous activity classification using Wi-Fi (Section \ref{simsense_architecture}). Unlike existing approaches, \FW can distinctly classify among different human subjects performing multiple activities simultaneously by utilizing multiple CSI collectors, each associated to a given subject;  \smallskip

$\bullet$ To address the challenge of generalizing to new environments and subjects, we propose a novel \gls{fsl}-based architecture called \FR. In stark contrast to existing approaches, \FR can adapt to any new environment and subject through two main steps, namely meta-learning and fine-tuning. Moreover, in contrast to the traditional \gls{fsl}, \FR combines both the embedding learning and meta-learning approaches to achieve better performance through fine-tuning the classifier with only a few additional samples (Section \ref{simsense_architecture});\smallskip

$\bullet$ We extensively evaluate \FW through an exhaustive data collection campaign in 3 different environments and with 3 subjects performing 20 different activities simultaneously. We demonstrate that \FW achieves classification accuracy of up to 98\%, while \FR improves the accuracy by 85\% in comparison to a traditional \gls{cnn} and up to 30\% when compared to the state of the art \gls{fsel} \cite{tian2020rethinking}, by only using 15 seconds of additional data for each class. 
\textbf{For reproducibility, we share our whole dataset, captured video streams of the activities as ground truth, and our code repository \cite{SiMWiSense}.}

\section{Sensing Proximity Test}\label{sec:prox_test}

Wi-Fi sensing leverages tiny changes in the \gls{csi} computed through pilot symbols included in the \gls{phy} preamble. Although the \gls{csi} may be captured by monitoring a transmission link between \gls{ap} and the \glspl{sta} without any direct communication with the AP, a monitoring device captures the \gls{csi} of the propagation channel between itself and the AP. Thus, when the \gls{csi} monitors are spatially distant enough, they would monitor the independent propagation path between the corresponding antenna pair of the AP and itself \cite{tse2005fundamentals}. Our key intuition is that the captured \gls{csi} is dominantly characterized by any physical change in the environment at spatially closer proximity.

\begin{figure}[h]
	\centering
	\includegraphics[width=\columnwidth]{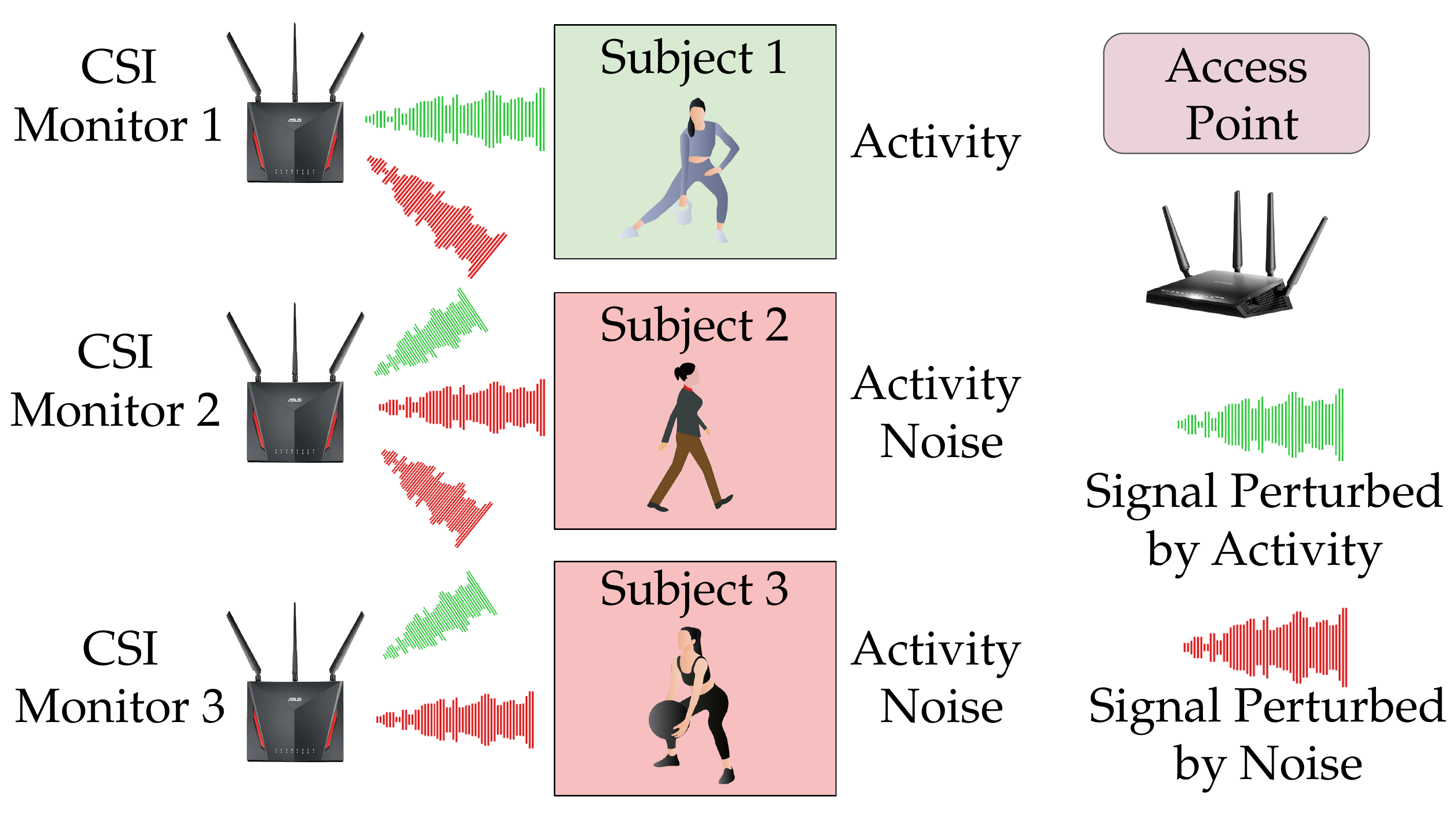}
	\caption{\FW proximity test.}
	\label{fig: proximity_simsense}
	\vspace{-0.2cm}
\end{figure}

To evaluate this, we perform the following extensive preliminary tests which demonstrate the viability of our key concept.  We have performed the sensing proximity test in 3 different environments with 3 different subjects and 20 activities. We assign a \gls{csi} monitor to each of the subjects. The monitors are placed at a distance of 1.5m - 3.0m from each other, whereas one human subject performs activity at a distance of 1.5m - 2.0m from each of the sensing monitors for each of the environments. We considered three different environments as explained in Section \ref{sec:exp_setup}. The experimental setup is shown in Figure \ref{fig: proximity_simsense}. We define the subjects with the closest proximity to the CSI Monitor 1, CSI Monitor 2 and CSI Monitor 3 as Subject 1, Subject 2 and Subject 3, respectively. From each environment, \gls{csi} is collected in three separate rounds where in every round a subject does 20 different activities and other subjects perform random activities. 

\begin{figure}[h]
	\centering
	\includegraphics[width=.93\columnwidth]{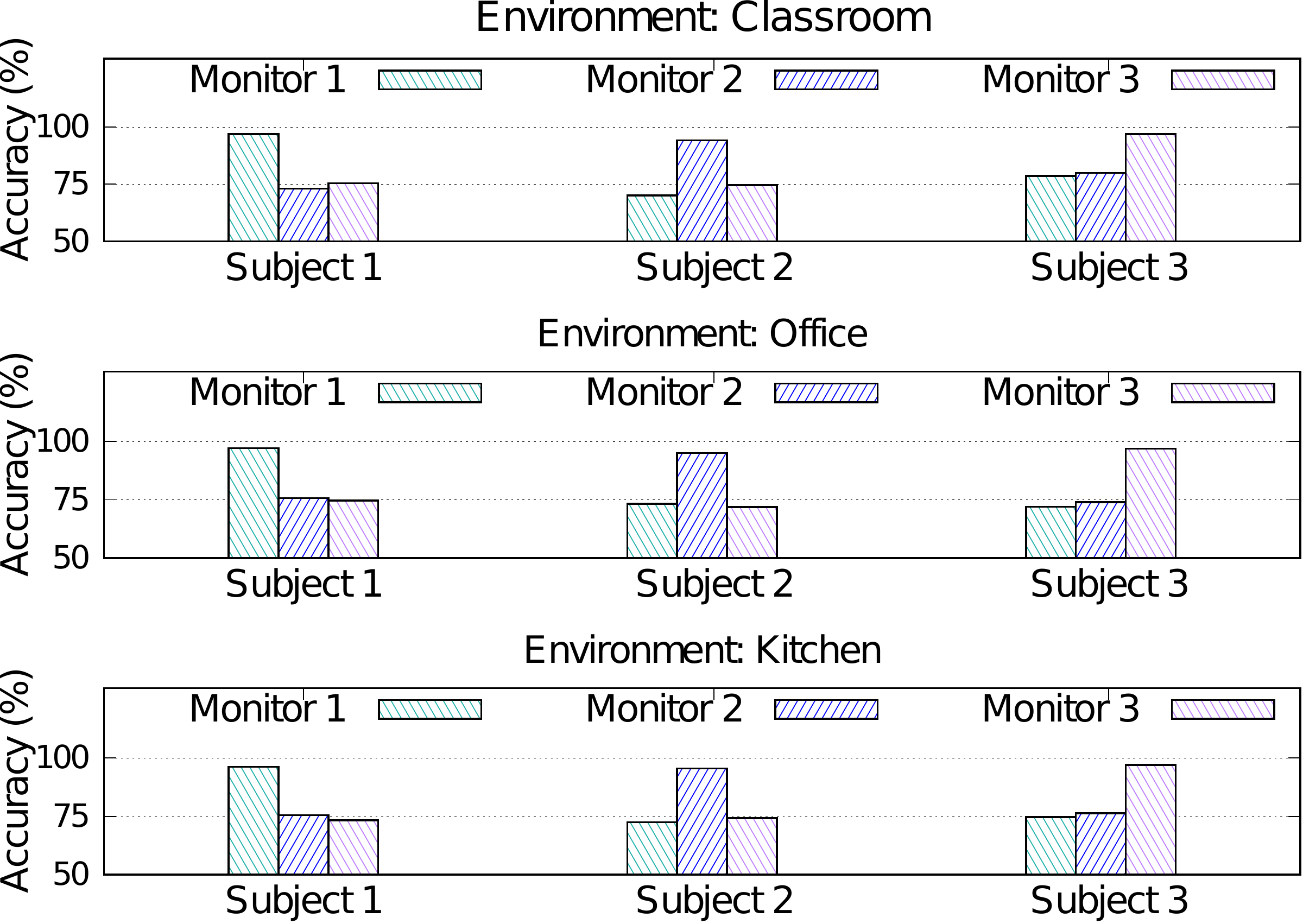}
	\caption{Classification accuracy as function of sensing proximity.}
	\label{fig:proximity}
\end{figure}

Figure \ref{fig:proximity} confirms our intuition. For example, it shows that in the classroom environment, the accuracy of Subject 1 is 95\% from the CSI data of Monitor 1 whereas, with the exact same setup and tests, the accuracy of Subject 2 and Subject 3 decreases by 30\% on an average with Monitor 1. This is because they are comparatively farther away from Subject 1 and more prone to the noises created by the other subjects at that instant. However, their performances improve drastically to 96\% and 97\% when we consider CSI Monitor 2 and CSI Monitor 3 for Subject 2 and Subject 3 respectively. The other two environments follow similar trends. \textbf{This clearly demonstrates that the CSI monitor closest to the subject performs better than other CSI monitors.} 

\vspace{-0.3cm}

\section{Overview of \FW}\label{simsense_architecture}

\begin{figure}[h!]
	\centering
	\includegraphics[width=0.95\columnwidth]{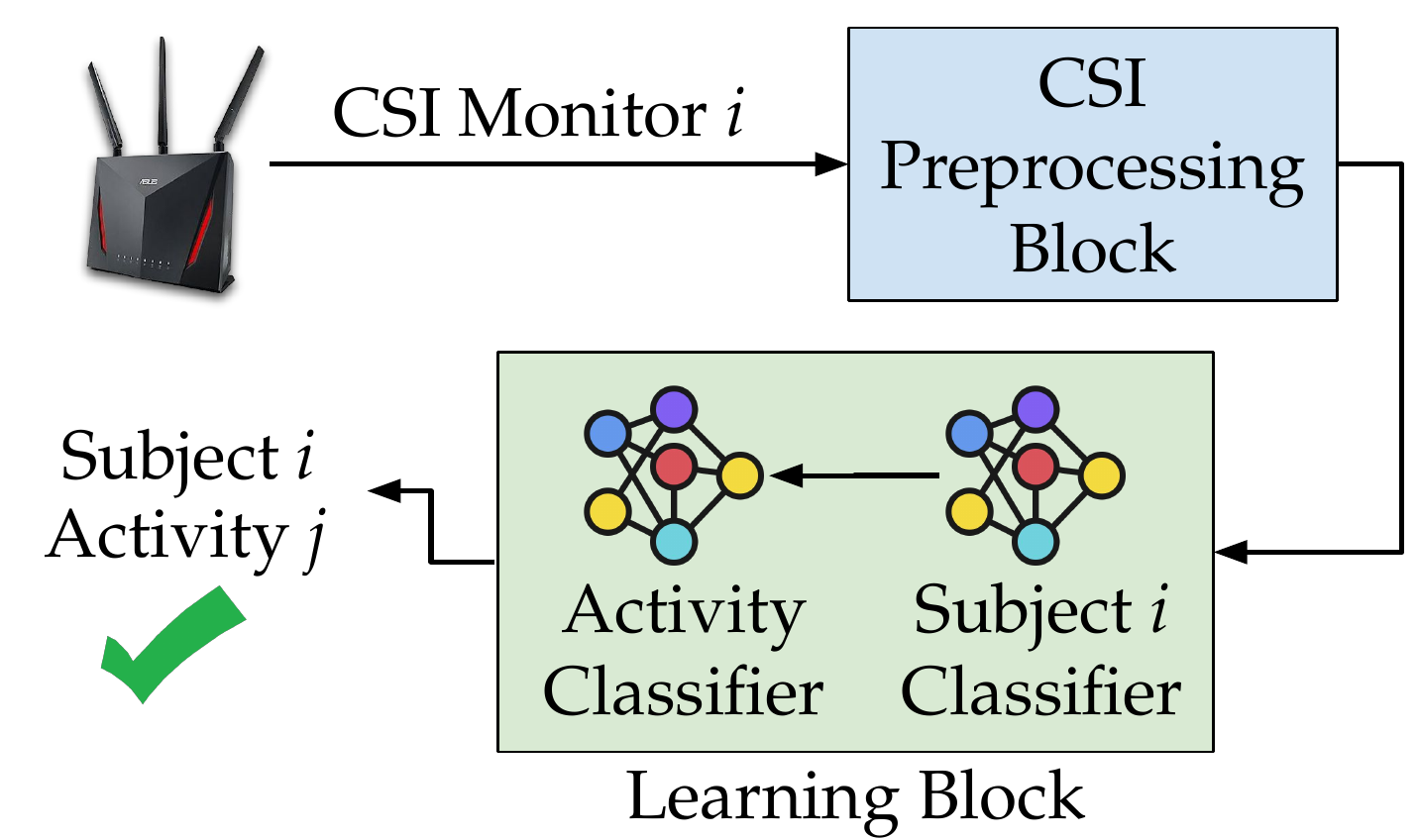}
	\caption{Overview of \FW and Processing Blocks.}
	\label{fig:SimSense_Architecture}
\end{figure}

We describe the \FW framework dividing it into three main task blocks: (i) sensing block (ii) preprocessing block and (iii) learning block as presented in Figure \ref{fig:SimSense_Architecture}.

\subsection{\FW Sensing and Preprocessing Blocks}

The sensing block of \FW collects the \gls{csi} of Wi-Fi transmissions. Modern Wi-Fi systems are based on the Orthogonal Frequency Division Multiplexing (OFDM) modulation which processes multiple data streams in parallel over multiple orthogonal subcarriers. Each spatially diverse \gls{csi} monitor captures $S$ samples during the time interval $T=t-t^{\prime}$ with $K$ orthogonal parallel subcarriers. Thus, the extracted \gls{csi} matrix of an $M\times N$ system is shown in Equation \ref{eqn:CFR_CSI}.

\begin{equation}
\label{eqn:CFR_CSI}
\begin{aligned}
H_r^{m,n} = 
\begin{bmatrix}
h_{1,1}^{m,n} & \dots & h_{1,s}^{m,n} & \dots & h_{1,K}^{m,n}\\
\vdots &   & \vdots & & \vdots \\
h_{s,1}^{m,n} & \dots & h_{s,s}^{m,n} & \dots  & h_{s,K}^{m,n}  \\
\vdots & \ddots & \vdots & \ddots & \vdots \\ 
h_{S,1}^{m,n} & \dots & h_{S,k}^{m,n} & \dots  & h_{S,K}^{m,n} 
\end{bmatrix}
\end{aligned}
\end{equation} 

Here, $H_r^{m,n}$ denotes the \gls{csi} matrix at receiver $r$ of the transmit antenna $m$ and receive antenna $n$ where  
$ 1 \leq n \leq N $ and $1 \leq m \leq M $. The value $h_{k,s}^{m,n}$ denotes the \gls{csi} of the S-th sample at K-th subcarrier from transmit antenna $m$ to the receive antenna $n$. For example, during the time interval $T=0.2s$, if any \gls{csi} monitor captures $S=600$ samples with a channel of 80 MHz bandwidth, $H_r^{m,n}$ will have $S\times K$ components where $K=242$. It is worthwhile mentioning that even though the total number of subcarriers in 80 MHz channel is 256, we only consider data-transmitting subcarriers discarding the null and the guard ones. 

\begin{figure}[h]
	\centering
	\includegraphics[width=0.95\columnwidth]{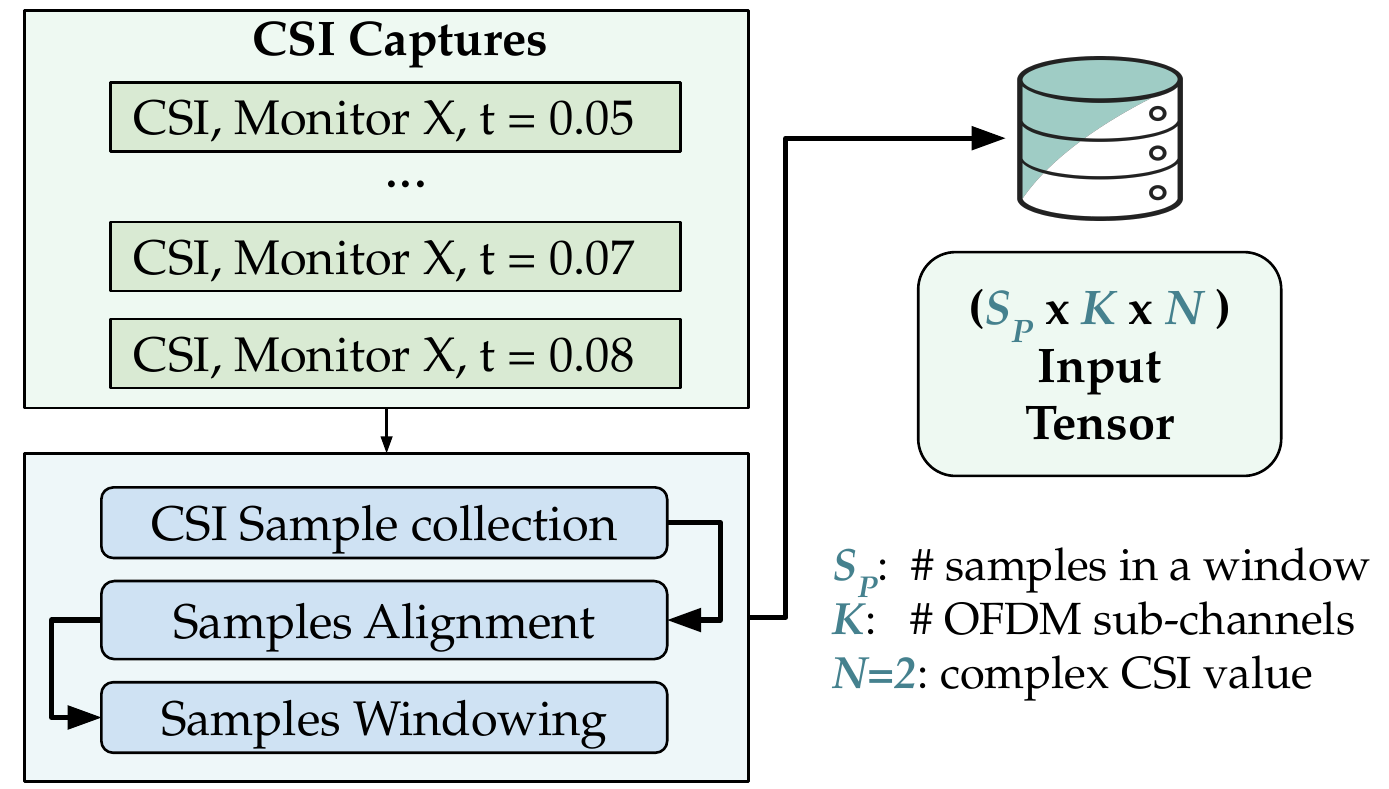}
	\caption{\gls{csi} data processing in \FW }
	\label{fig:csi_preprocess} 
		\vspace{-0.2cm}
\end{figure}

After the collection of the \gls{csi} samples, we preprocess the captured data, as presented in Figure \ref{fig:csi_preprocess}, before it is fed to the learning block.  After $S$ samples are collected, we align the data by discarding the  missing and/or corrupted \gls{csi} measurements. Moreover, any abrupt amplification in the data is removed by normalizing with the mean \gls{csi} amplitude. Then, the captures are segmented with a fixed size non-overlapping window along the time domain. If the total number of samples captured during any time interval $T$ is $S$, such that $T = T_1 + T_2 + T_3 +.......+ T_n $ where $T$ is divided into $n$ equal time windows, $S= S_1 + S_2 + S_3 +........+ S_n$ are the corresponding sample captures of the $n$ time segments. Thus, each window has the tensor dimension of $S_p\times K\times N$ where $S_p$ is the number of samples in p-th time window $T_p$, and $N=2$ is the complex \gls{csi} measurement. This processed data is then fed to the input of the learning block. 

\subsection{\FW Learning Block}

One of the challenges in multi-subject detection is scalability. For $P$ persons and $Q$ activities, it has $Q^P$ possible combinations, resulting in an exponential increase in the number of classes. One centralized model to classify multi-subject activities becomes difficult when $P$ and $Q$ are large. To tackle this problem, we propose a decentralized detection system for each subject. Specifically, a learning model is assigned to each device to sense the subject which is closest to it. Therefore, each subject only requires $Q$ detection regions in hyperspace. For $P$ subjects, the overall complexity reduces to $P \times Q$. This sensing system has two assumptions: (i) there will be at least the same number of CSI collectors as subjects; (ii) the subject closest to the device take the most significant part in shaping the channel property between the device and \ap. Assumption (i) is reasonable since nowadays, almost everyone is inseparable from their smart devices, such as laptops or smartphones, in their work and daily lives. For (ii), we developed an experiment based on sensing proximity as shown in Section \ref{sec:prox_test}.

\begin{figure}[ht]
	\centering
	\includegraphics[width=.48\textwidth]{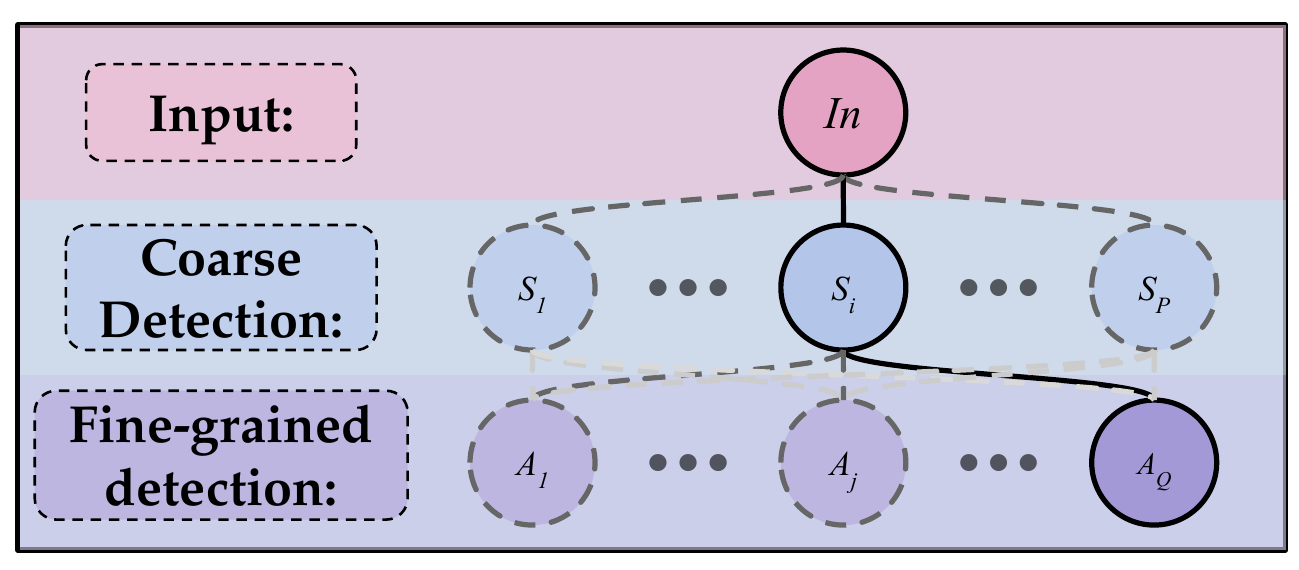}
	\caption{Cascaded Learning Block.\vspace{-0.2cm}}
	\label{fig:learningblock}
\end{figure}

To further decrease complexity, we propose a cascaded model for two-stage detection. As shown in Figure \ref{fig:learningblock}, in the first stage, a \gls{dl} model is used to discriminate different subjects $S$ coarsely. After the coarse detection, another fine-grained \gls{dl} model is used to determine the activities $A$. Regardless of the output at the first stage, all the subjects will share the same fine-grained model at the second stage. Thus, the overall complexity becomes $P+Q$.

One challenging problem of the hierarchical detection model is that even if different persons do the same activities, their movements will have personal patterns and gestures. Furthermore,  subjects may join or leave the detection system. Thus, it is impractical to have a universal classifier for activity detection. In addition, the performance of data-driven algorithms in wireless sensing usually will be downgraded by time-varying channel conditions.  Thus, we need a model which can swiftly adapt to new subjects and channel features.


\subsection{\FR \space Learning Algorithm}

We propose a novel \gls{fsl} architecture named \FR, which allows the \gls{dl} model to adapt to new scenarios with only a few data. We utilize this algorithm in both subject detection and activity detection stages. Next, we will discuss the algorithm in detail.  \gls{fsl} aims at training models that can rapidly generalize to new tasks with only a limited number of labeled samples, is a strong candidate to tackle the data collection problem. One approach to FSL is to learn an embedding for multiple tasks \cite{vinyals2016matching,snell2017prototypical}. Specifically, a \gls{dnn} is used to learn a clustered mapping from input to the latent space. During the inference time, the embedding network does not need to be fine-tuned, and a few samples will be used as references to classify unobserved data. Another approach is meta-learning \cite{finn2017model,nichol2018first} which involves two phases: (i) meta-training and (ii) fine-tuning. Meta-learning aims to learn shared features between different tasks during the meta-learning phase and quickly optimize the parameters with a few data points during the fine-tuning stage.

\begin{figure}[ht]
	\centering
	\includegraphics[width=.48\textwidth]{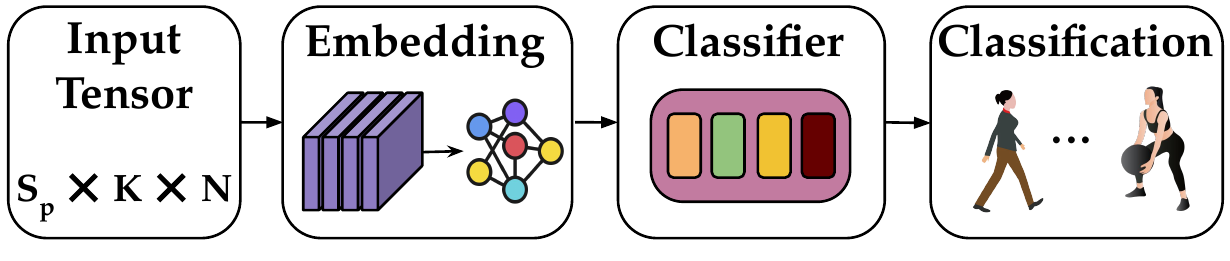}
	\caption{Feature Reusable Embedding Learning (FREL).}
	\label{fig:free}
\end{figure}

For the first time, \FR combines embedding learning and meta-learning. Figure \ref{fig:free} demonstrates the structure of our \gls{frel} model. First, a \gls{dnn} model is used to learn the embedding of the input and another classifier is used to decode the features in the latent space. Similar to meta-learning, the \gls{frel} consists of meta-learning and fine-tuning stages. We train the embedding network and classifier jointly with a mini-dataset during the meta-learning phase, which is the same as embedding learning. After obtaining the embedding, we further optimize the classifier with a few samples while testing. In contrast to embedding learning, we believe fine-tuning can provide better flexibility and granularity, which is more suitable for a dynamic system. Different from meta-learning, we only retrain the simple classifier instead of the whole structure, which can reduce the computation, enabling a faster adaption to new tasks. This design is inspired by \cite{raghu2019rapid}, which shows that the effectiveness of meta-learning is mainly due to feature reuse. It is a simplified version of MAML \cite{finn2017model} that fine-tunes the last few layers which can achieve comparable performance as the original algorithm.

Formally, we consider the embedding network as a function $E_\theta:X\rightarrow Z$, where $Z$ denotes the latent vector. The classifier $C_\phi:Z\rightarrow Y$ is to find a mapping between encoded features $Z$ and labels $Y$. $\theta$ and $\phi$ are the trainable parameters of the embedding network and classifier, respectively. Hence, The overall system $F_\psi(X) = Y$ can be written as $C_\phi(E_\theta(X))=Y$, where $\psi=\{\theta,\phi\}$ is the total trainable parameters of the whole system. In \gls{frel}, models are trained on a set of mini-batches of data that only have N different classes (ways) and K samples (shots) of each class. Each batch of few-shot data can be considered as a new task $\tau_j=\{(x^{j}_{i}, y^{j}_{i})\}|^{m}_{i=1}$ in meta-learning. $m=N \times K$ denotes the total number of samples in one batch. The objective of meta-learning is to find a set of parameters $\psi$ that minimize the expectation of the loss function $\mathcal{L}$ with respect to a group of meta-learning tasks $\mathcal{T} = \{\tau_j\}|^{n}_{j=1}$, i.e.,

\begin{equation}
    \begin{aligned}
        \label{eqn:free1}
        \min_{\{\theta, \phi\}}&&\frac{1}{n}\sum_{j=1}^{n} [ \frac{1}{m}\sum_{i=1}^{m} \mathcal{L}(C_\phi(E_\theta(x_i^j)) = y_i^j)]
    \end{aligned}
\end{equation}

We merge the task set $\mathcal{T}$ into a single dataset $\mathbb{D}^{train}$ to get a better embedding, which is given by

\begin{equation}
    \begin{aligned}
        \label{eqn:free2}
        \mathbb{D}^{train} & = \tau_1 \cup \cdots \cup \tau_j \cup \cdots \cup \tau_n\\
        & = \{(x^{1}_{i}, y^{1}_{i})\}|^{m}_{i=1} \cup \cdots \cup \{(x^{n}_{i}, y^{n}_{i})\}|^{m}_{i=1}
    \end{aligned}
\end{equation}

We notice that by merging multiple tasks into single dataset, the optimization problem in Equation \ref{eqn:free1} can be reduced to a general \gls{dl} problem, which can be solved by a gradient decent optimizer iteratively,
\begin{equation}
    \begin{aligned}
        \label{eqn:free3}
        \{\theta,\phi\} = \{\theta,\phi\} - \alpha \frac{1}{mn} \sum_{i=1}^{mn} \nabla_{\{\theta,\phi\}}\mathcal{L}(C_\phi(E_\theta(x_i)),y_i)
    \end{aligned}
\end{equation}
where $mn$ is the total number of data points in $\mathbb{D}^{train}$ and $\alpha$ is the learning rate. Once the optimal embedding $\theta^\ast$ is obtained, the classifier is fine-tuned on another small portion of data $\mathbb{D}^{tune}$. Unlike training on a combined set during the meta-learning, each iteration we randomly sample K shots from each of N ways in $\mathbb{D}^{tune}$ to build a new task $\tau$ and update classifier by gradient decent,
\begin{equation}
    \begin{aligned}
        \label{eqn:free4}
        \phi = \phi - \beta \frac{1}{m} \sum_{i=1}^{m} \nabla_{\phi}\mathcal{L}(C_\phi(E_{\theta^\ast}(x_i)),y_i)
    \end{aligned}
\end{equation}
where $\beta$ denotes the learning rate in the fine-tuning phase. Finally, performance is evaluated on the rest of unseen dataset $\mathbb{D}^{test}$. 

We summarize \gls{frel} in Algorithm \ref{algo:free}. It is worth pointing out that although \gls{frel} is proposed for WiFi sensing initially, the architecture is presented generally since it can be used for other \gls{fsl} purposes. Next we discuss the specific setup that is used in the \FW learning block.\smallskip

\begin{figure}[ht]
	\centering
	\includegraphics[width=0.95\columnwidth]{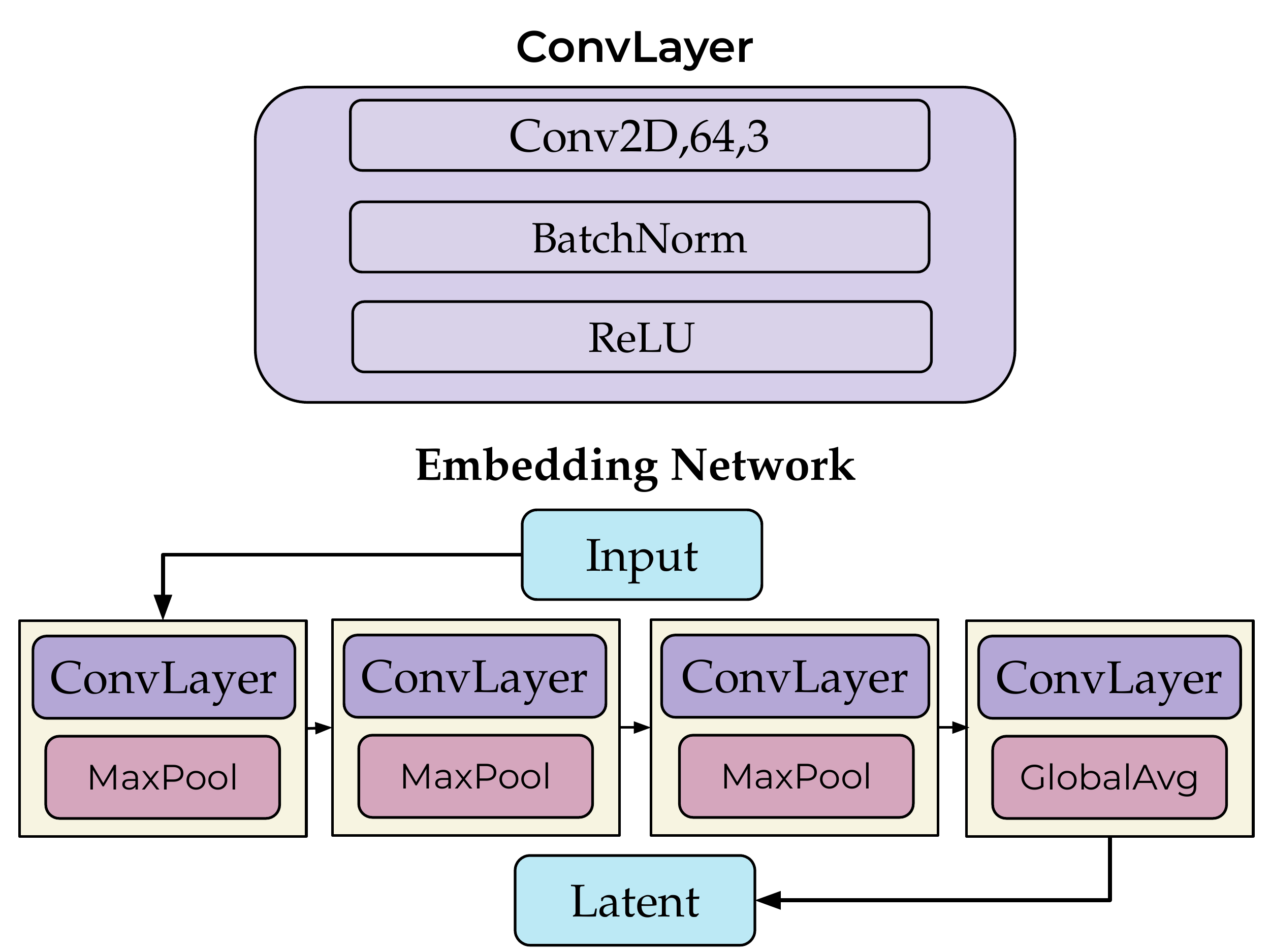}
	\caption{CNN Embedding Network.}
	\label{fig:SimSense_CNN}
\end{figure}

$\bullet$ \textbf{Embedding network:} Figure \ref{fig:SimSense_CNN} shows the \gls{dnn} architecture we use for embedding learning in \gls{frel}. It is composed of 4 convolutional layers followed by batch normalization and \gls{relu} activation. Each convolutional layer comprises 64 channels with a kernel size of $3\times3$. After the first three convolutional layers, $2\times2$ Max pooling layers are used to down sample the previous layer's output. After the fourth convolutional layer, a global average pooling strategy is chosen to extract the feature to the latent space, resulting in a 64-dimensional feature space.

\RestyleAlgo{ruled}
\SetAlgoNoLine
\LinesNotNumbered
\begin{algorithm}
\caption{Feature Reusable Embedding Learning} \label{algo:free}
\texttt{\\}
\textbf{\textit{Phase 1: \gls{frel} meta-learning}} \\
Require: learning rate $\alpha$, dataset $\mathbb{D}^{train}$\\
Initialize: $\theta$ for embedding, $\phi$ for classifier\\
\For{$iteration=1,2,...$}{
update $\theta$ and $\phi$ with $\mathbb{D}^{train}$ by Equation \ref{eqn:free3} 
}
Return: $\theta^\ast$ for embedding\\
\texttt{\\}
\textbf{\textit{Phase 2: \gls{frel} fine-tuning}}\\
Require: learning rate $\beta$, dataset $\mathbb{D}^{tune}$\\
Initialize: $\phi$ for classifier\\
\For{$epoch=1,2,...$}{
\For{$episode=1,2,...$}{
sample a task $\tau=\{(x_i,y_i)|_{i=1}^{m}\}$ from $\mathbb{D}^{tune}$\\
update $\phi$ with $\tau$ by Equation \ref{eqn:free4}
}
}
Return: $\phi^\ast$ for classifier
\end{algorithm}

$\bullet$ \textbf{Classifier:} A fully-connected layer is used on top of the pre-trained embedding network as a linear decoder. Non-linearity functions such as \gls{relu} are not applied since we aim to study the efficacy of the overall \gls{frel}'s design rather than develop complicated \gls{dnn} models. To investigate the effectiveness of fine-tuning in \gls{frel}, we also implement an untrainable \gls{knn} algorithm as a comparison after the meta-learning phase following the same procedure as another state-of-the-art \gls{fsel} model\cite{tian2020rethinking}. During the inference time of \gls{knn}, K samples from each class are transformed into embedding as supports, and the queries are classified by a plurality vote of the K nearest supports.\smallskip

$\bullet$ \textbf{Mini-dataset:} Usually in general \gls{fsl}, the dataset such as Omniglot \cite{lake2011one} and Mini-ImageNet \cite{vinyals2016matching} contains a large number of tasks and a few number of samples in each task. Algorithms are first pre-trained on multiple tasks and then fine-tuned and tested on single specific task. However, it is never feasible to get a dataset with comprehensive tasks in wireless sensing since the changing environment can always generate new tasks that models have never seen before. Thus, one significant difference in our implementation is the mini-dataset. We only utilize limited number of data collected in 15 seconds for $\mathbb{D}^{tune}$. The test set $\mathbb{D}^{test}$ is never exposed to the model during pre-training and fine-tuning. The mini-dataset makes the problem more challenging as models are learned from not only a few samples but a few tasks.\smallskip

$\bullet$ \textbf{Learning strategy:} We evaluate our model with 5-shot learning, which means we have 5 samples for each class in every mini-batch data. We choose Adam as the optimizer in both phases. The learning rate $\alpha$ and $\beta$ are 0.01. Cross-entropy loss is used during the pre-training and fine-tuning stages for simplicity. Other metrics such as deep k-means \cite{fard2020deep} and prototypical loss \cite{snell2017prototypical} can be applied for different purpose.

\vspace{-0.1cm}

\section{Experimental Evaluation} \label{experimental_setup_and_performance_evaluation}

\subsection{Experimental Setup}\label{sec:exp_setup}

We evaluate the multi-subject sensing capability of \FW as well as the generalizing feature of \FR through an extensive data collection campaign in 3 different environments: classroom, office and kitchen with 3 human subjects performing 20 different activities in random order. We used off-the-shelf Netgear Nighthawk X4S AC2600 routers to set up the network, whereas IEEE 802.11ac compliant Asus RT-AC86U routers with Nexmon tool are used as the \gls{csi} extractor\cite{nexmoncsi2019}. 

\begin{figure}[ht]
	\centering
	\includegraphics[width=.49\textwidth]{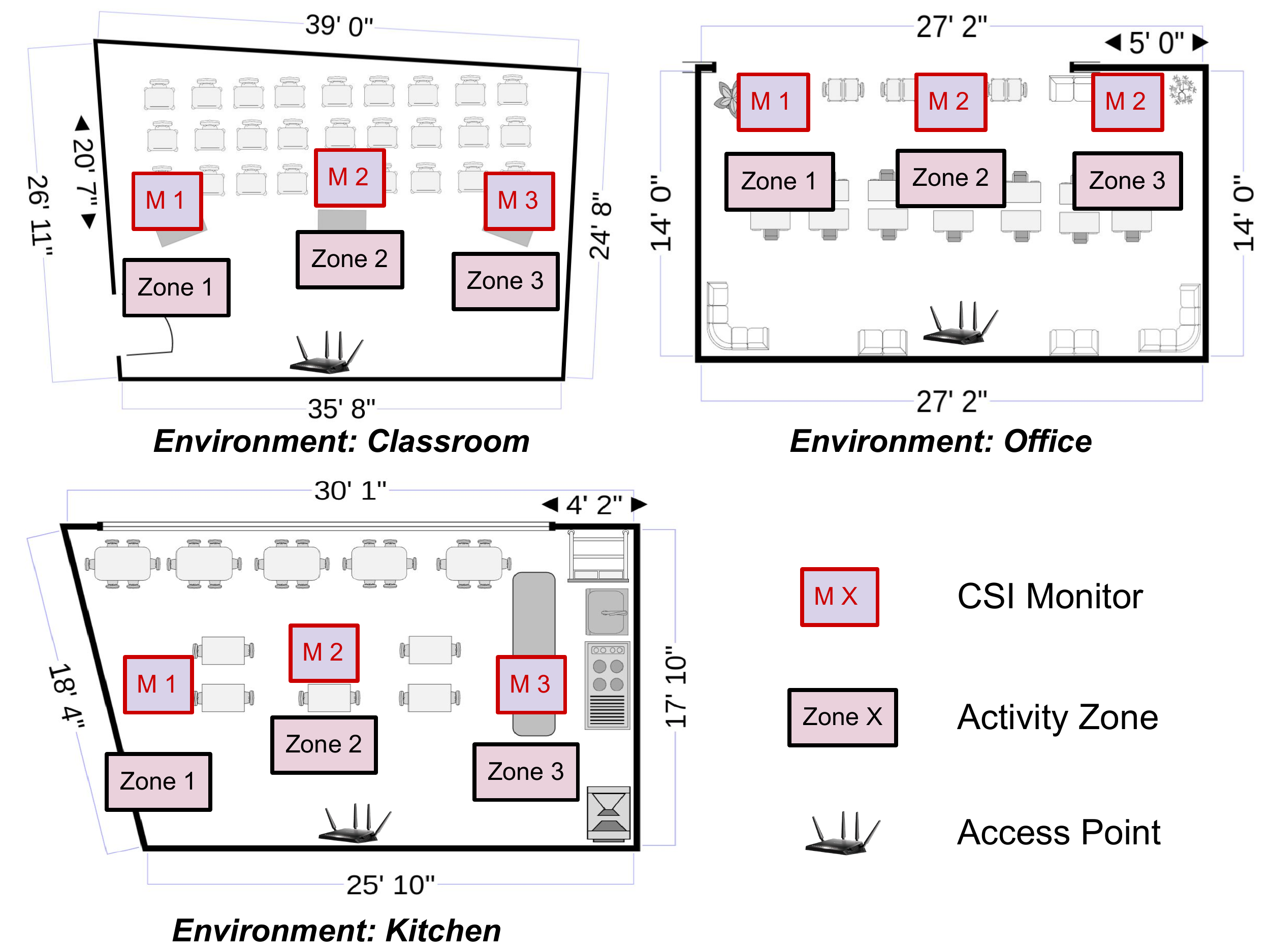}
	\caption{Experimental Setup of \FW.}
	\label{fig:environments}
\end{figure}

The AP and the \gls{sta}s are configured with $M=3$ antennas and $N_{ss}=3$ spatial streams, while the \gls{csi} monitors are configured with $M=1$ antenna and $N_{ss}=1$ spatial streams respectively. We send UDP packets from AP to the STAs to trigger the \gls{csi} monitors. The \gls{csi} has been collected at center frequency $f_c=5.21 GHz$  (i.e., channel 42) with signals having 80 MHz bandwidth. Three \gls{csi} monitors were placed in each of the environments at a distance of 1.5m - 2.0m from each other as shown in Figure \ref{fig:environments}. Three different subjects performed all the 20 activities: \textit{push forward, rotate, hands up and down, waive, brush, clap, sit, eat, drink, kick, bend forward, wash hands, call, browsing phone, check wrist, read, waive while sitting, writing, side bend, and standing} at a distance of 1.5m -2.0m from the \gls{csi} monitors.

\begin{figure}[h]
	\centering
	\includegraphics[width=.49\textwidth]{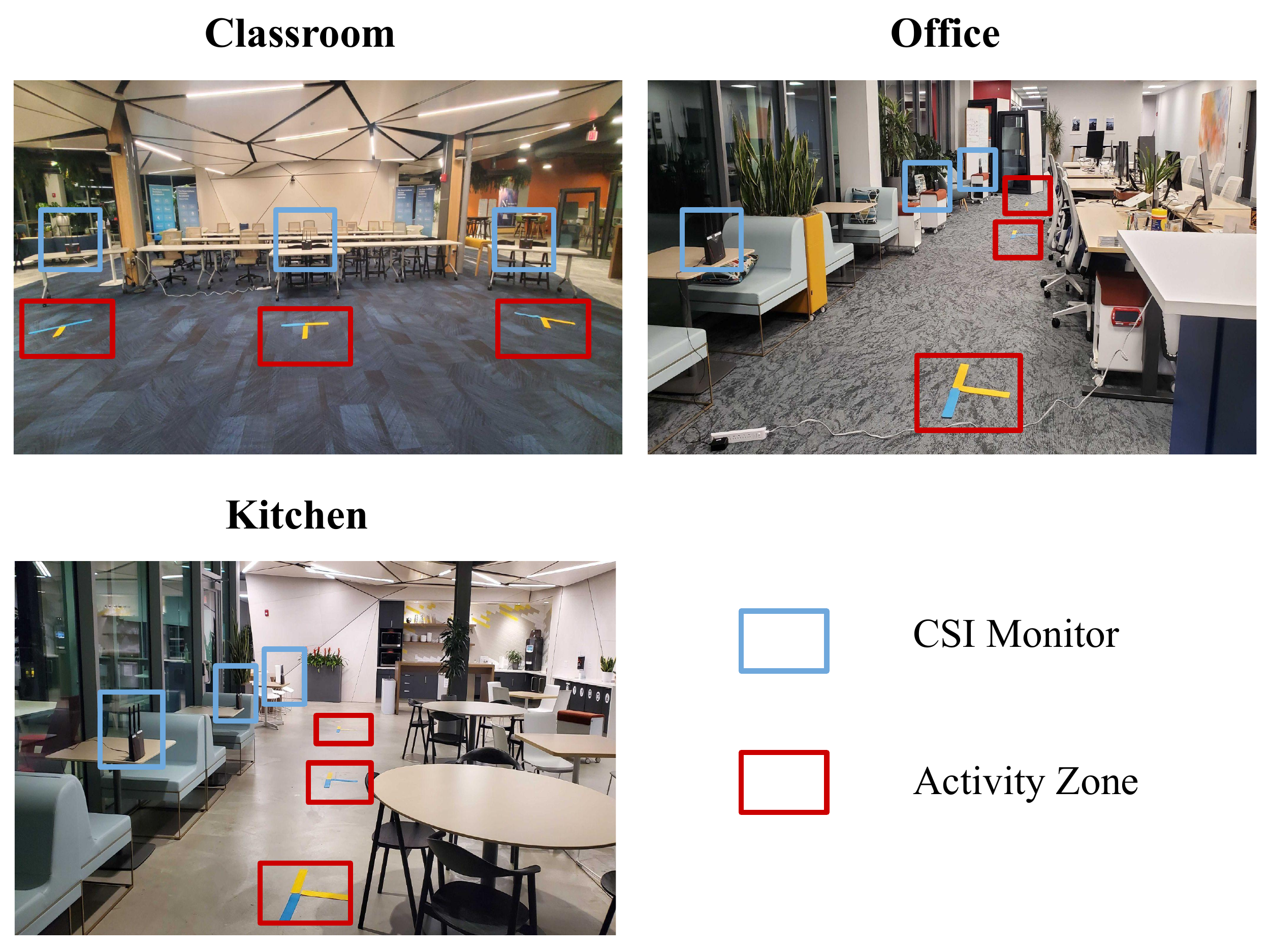}
	\caption{Data collection locations.\vspace{-0.3cm}}
	\label{fig:environmental_site}
\end{figure}

Three separate models are trained for each of the monitors with the data of each corresponding subject performing the activities while the other subjects do random activities. For example, the model for Monitor 1 is trained with Subject 1 performing each of the 20 activities for 5 minutes while Subject 2 and Subject 3 perform different random activities in random order. Similarly, the data from each of the monitors are collected at the mentioned three different environments and for each subject. To evaluate the simultaneous multi-subject sensing, testing data has been collected while different subjects perform randomly chosen activities simultaneously. The experimental setup and the activity zone of the subjects at the mentioned three experimental sites are presented in Figure \ref{fig:environmental_site}. To create the ground truth, we captured the video streams of the subjects performing different activities in synchronization with the data collection.

\subsection{Performance Evaluation of \FW} \label{performance_evaluation}
A time window size of 0.1s is considered for data segmentation, while each data window has 50 samples for all the tests performed in \ref{performance_evaluation}. Firstly, we do the performance evaluation of our two-stage detection system: (i) subject identification (ii) activity classification with baseline \gls{cnn} as shown in Figure \ref{fig:SimSense_CNN}. Then, we demonstrate the generalization capability of \FR in both of the stages of the detection system and compare the performance of \FR with both baseline \gls{cnn} and state-of-the-art \gls{fsel} \cite{tian2020rethinking} model. In the rest of \ref{performance_evaluation}, Monitor 1, Monitor 2 and Monitor 3 are denoted as M1, M2, and M3 respectively whereas Subject 1, Subject 2 and Subject 3 are presented as Sub1, Sub2 and Sub3 respectively.\smallskip

\begin{figure}[ht]
	\centering
	\includegraphics[width=.93\columnwidth]{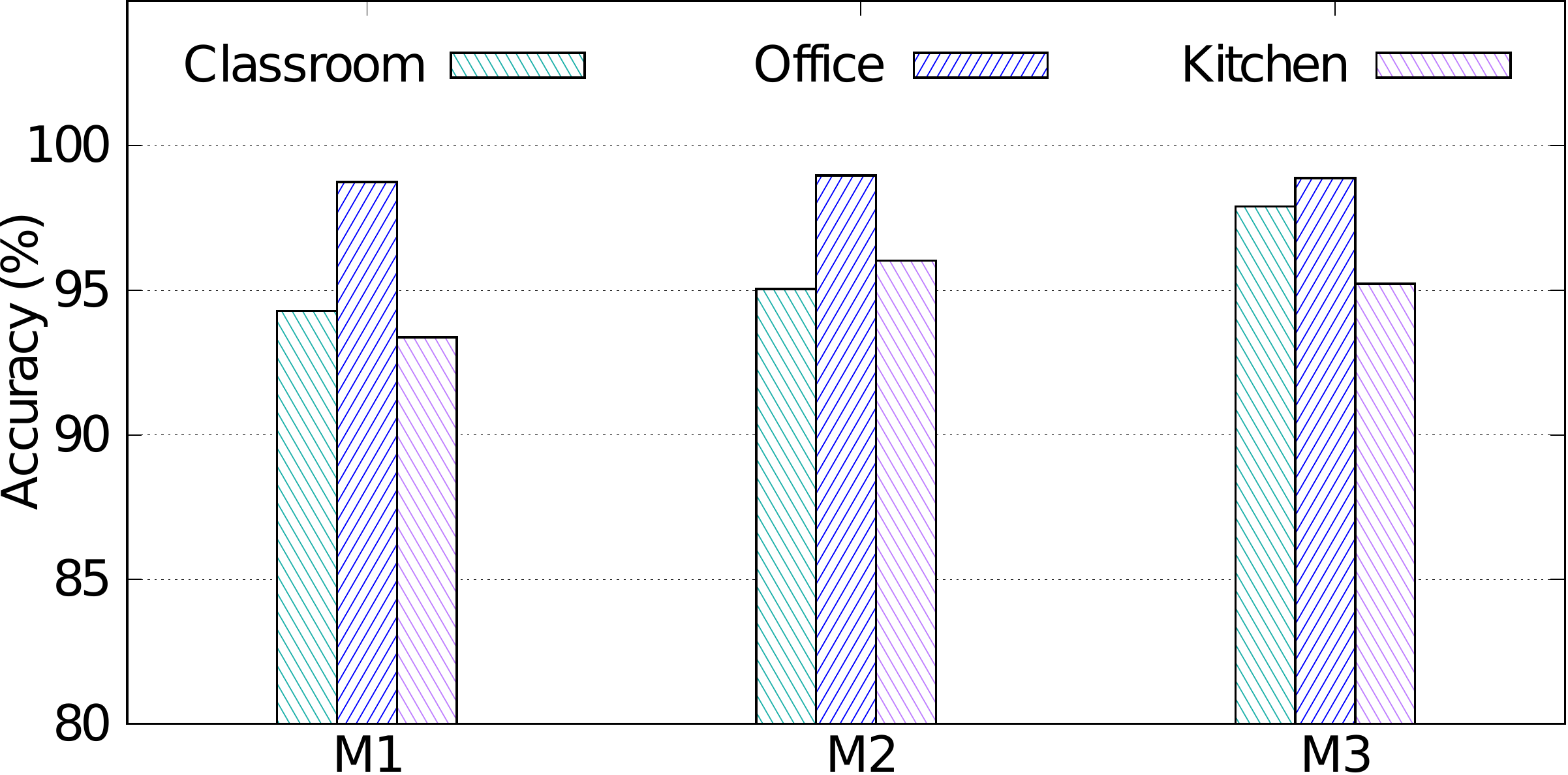}
	\caption{Subject identification in \FW with baseline CNN}
	\label{fig:Anomaly}
\end{figure}

\textbf{Baseline \gls{cnn}:} In the first step of the two-stage \FW detection, each \gls{csi} monitor classify Sub1, Sub2, Sub3 or "no activity". The subject identification performances of each of the monitors in three environments are presented in Figure \ref{fig:Anomaly}.

The results show that with the baseline \gls{cnn}, the average accuracy of M1, M2 and M3 across all three environments are 95.47\%, 96.68\% and 97.34\% respectively, which shows no significant performance discrepancy. On the other hand, the average performances in three different environments are 95.73\%, 98.77\% and 94.87\% respectively which shows an average of 3.51\% performance boost in office compared to the other environments. It is caused by the Non-line-of-sight propagation between the monitors and the distant subjects due to the presence of desks and computers, causing less noise in identifying the closest subject.

\begin{figure}[ht]
	\centering
	\includegraphics[width=.93\columnwidth]{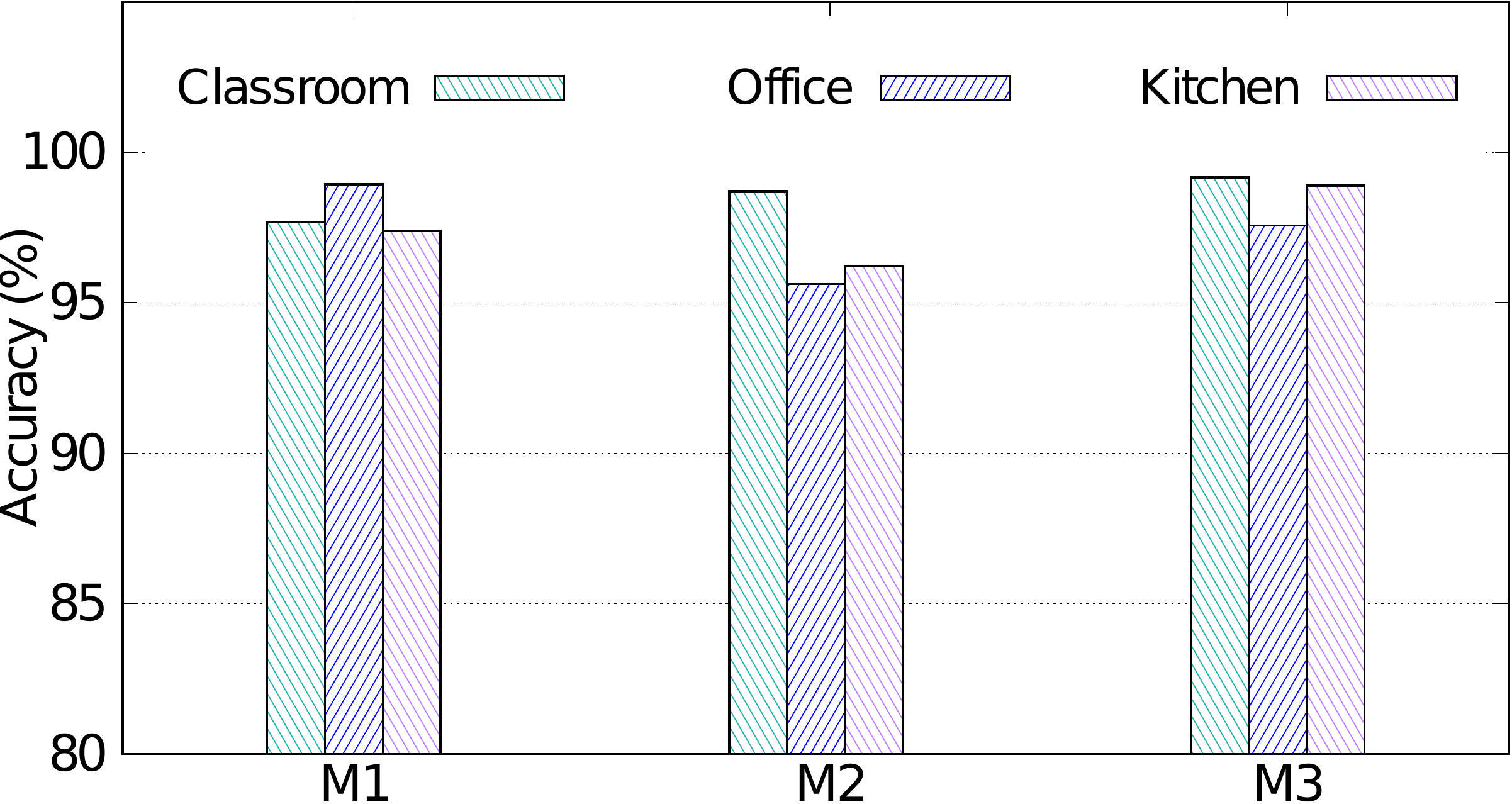}
	\caption{Performance of \FW at three different environments with baseline CNN as presented in Figure \ref{fig:SimSense_CNN}}
	\label{fig:classification_cnn}
\end{figure}

The simultaneous multi-subject activity classification performance of \FW with baseline \gls{cnn} is presented in Figure \ref{fig:classification_cnn}. The average accuracy in the environments: office, classroom and kitchen are 98.51\%,  97.37\% and 97.49\% respectively which follows the similar trend in performance depicting the stability and robustness of \FW. Moreover, the performance discrepancy of monitors M1, M2 and M3 are less than 2\% achieving an average accuracy of  98.0\%, 96.84\% and 98.54\% respectively with M1, M2 and M3. \smallskip

\textbf{Performance as a function of subcarrier resolution:} It is known that Wi-Fi sensing performs worse with lower subcarrier resolutions \cite{shi2019synthesizing, zeng2020multisense}. To compensate the lower subcarrier resolution, one can adopt extensive feature extraction techniques or higher sampling frequency which would increase the computation burden by intensifying the pre-processing steps and learning process dramatically. This stimulates us to study the trade-off of the subcarrier resolution and the \FW performance.  

\begin{figure}[ht]
	\centering
	\includegraphics[width=.93\columnwidth]{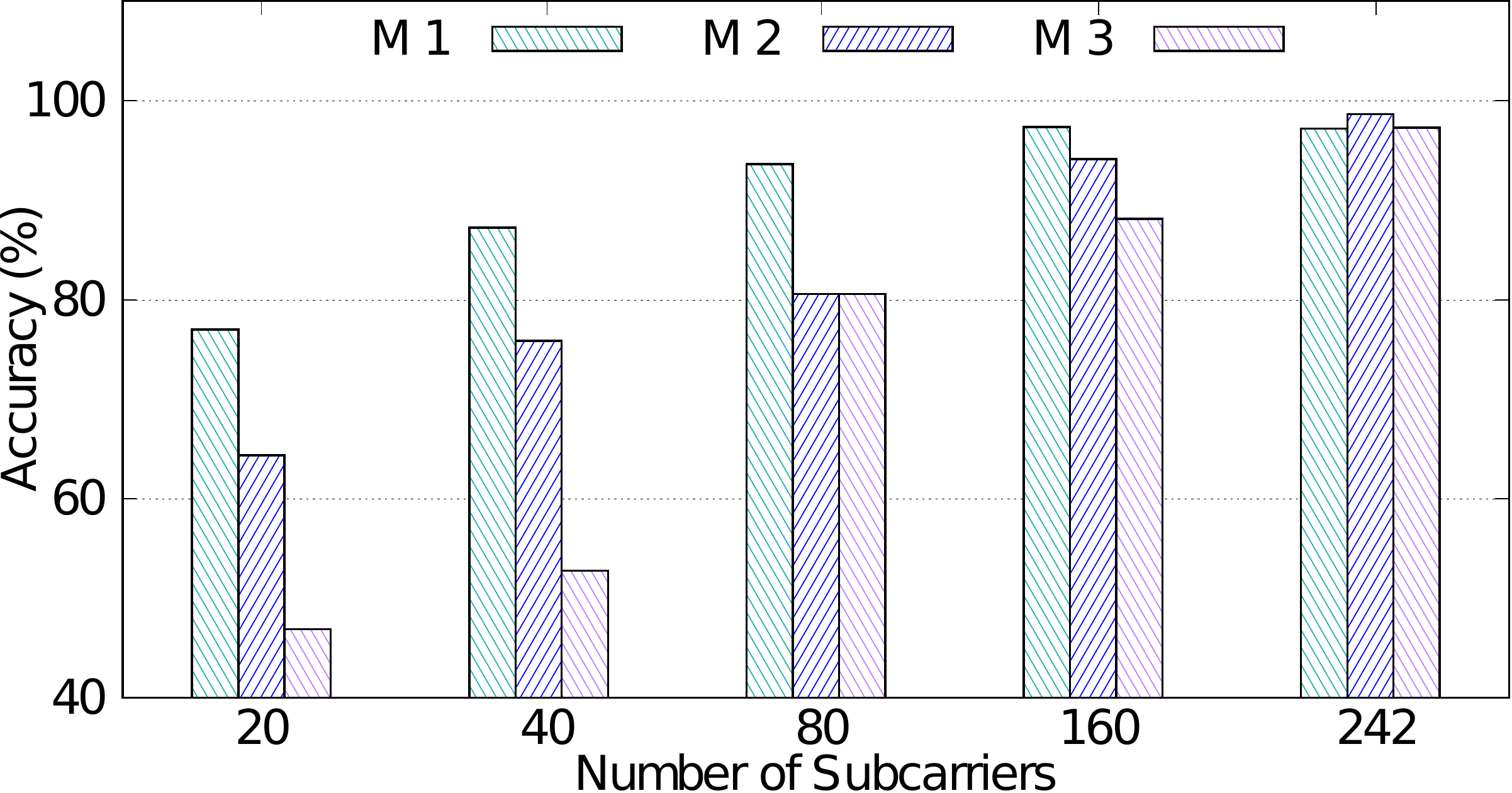}
	\caption{Performance of \FW with baseline \gls{cnn} as a function of number of subcarriers (environment: classroom). }
	\label{fig:subcarrier_resolution}
\end{figure}

\begin{figure}[b]
    \centering
    \begin{subfigure}[t]{0.49\columnwidth}
    \centering
    \setlength\fwidth{.7\columnwidth}
    \setlength\fheight{0.6\columnwidth}
    \begin{tikzpicture}
\pgfplotsset{every tick label/.append style={font=\tiny}}

\begin{axis}[
enlargelimits=false,
colorbar,
colormap/Purples,
width=\fwidth,
height=\fheight,
at={(0\fwidth,0\fheight)},
scale only axis,
tick align=inside,
xlabel={Predicted Activity},
xmin=-0.5,
xmax=19.5,
xtick style={draw=none},
xlabel style={font=\scriptsize\color{white!15!black}},
ylabel style={font=\scriptsize\color{white!15!black}},
ylabel={Actual Activity},
ymin=-0.5,
ymax=19.5,
xlabel shift=-5pt,
ylabel shift=-5pt,
ytick style={draw=none},
axis background/.style={fill=white},
colorbar horizontal,
colorbar style={
at={(0,1.05)},               
anchor=below south west,    
width=\pgfkeysvalueof{/pgfplots/parent axis width},
xtick={0, 0.5, 1},
xmin=0,
xmax=1,
axis x line*=top,
xticklabel shift=-1pt,
point meta min=0,
point meta max=1,
},
colorbar/width=2mm,
]
\addplot [matrix plot,point meta=explicit]
 coordinates {
(0,0) [1] (0,1) [0.0] (0,2) [0.0] (0,3) [0.0] (0,4) [0.0] (0,5) [0.0] (0,6) [0.0] (0,7) [0.0] (0,8) [0.0] (0,9) [0.0] (0,10) [0.0] (0,11) [0.0] (0,12) [0.0] (0,13) [0.0] (0,14) [0.0] (0,15) [0.0] (0,16) [0.0] (0,17) [0.0] (0,18) [0.0] (0,19) [0.0]

(1,0) [0.0] (1,1) [1.0] (1,2) [0.0] (1,3) [0.0] (1,4) [0.0] (1,5) [0.0] (1,6) [0.0] (1,7) [0.0] (1,8) [0.0] (1,9) [0.0] (1,10) [0.0] (1,11) [0.0] (1,12) [0.0] (1,13) [0.0] (1,14) [0.0] (1,15) [0.0] (1,16) [0.0] (1,17) [0.0] (1,18) [0.0] (1,19) [0.0]

(2,0) [0.0] (2,1) [0.0] (2,2) [1.0] (2,3) [0.0] (2,4) [0.0] (2,5) [0.0] (2,6) [0.0] (2,7) [0.0] (2,8) [0.0] (2,9) [0.0] (2,10) [0.0] (2,11) [0.0] (2,12) [0.0] (2,13) [0.0] (2,14) [0.0] (2,15) [0.0] (2,16) [0.0] (2,17) [0.0] (2,18) [0.0] (2,19) [0.0]

(3,0) [0.0] (3,1) [0.0] (3,2) [0.0] (3,3) [1.0] (3,4) [0.0] (3,5) [0.0] (3,6) [0.0] (3,7) [0.0] (3,8) [0.0] (3,9) [0.0] (3,10) [0.0] (3,11) [0.0] (3,12) [0.0] (3,13) [0.0] (3,14) [0.0] (3,15) [0.0] (3,16) [0.0] (3,17) [0.0] (3,18) [0.0] (3,19) [0.0]

(4,0) [0.0] (4,1) [0.0] (4,2) [0.0] (4,3) [0.0] (4,4) [1.0] (4,5) [0.0] (4,6) [0.0] (4,7) [0.0] (4,8) [0.0] (4,9) [0.0] (4,10) [0.0] (4,11) [0.0] (4,12) [0.0] (4,13) [0.0] (4,14) [0.0] (4,15) [0.0] (4,16) [0.0] (4,17) [0.0] (4,18) [0.0] (4,19) [0.0]

(5,0) [0.0] (5,1) [0.0] (5,2) [0.0] (5,3) [0.0] (5,4) [0.0] (5,5) [1.0] (5,6) [0.0] (5,7) [0.0] (5,8) [0.0] (5,9) [0.0] (5,10) [0.0] (5,11) [0.0] (5,12) [0.0] (5,13) [0.0] (5,14) [0.0] (5,15) [0.0] (5,16) [0.0] (5,17) [0.0] (5,18) [0.0] (5,19) [0.0]

(6,0) [0.0] (6,1) [0.0] (6,2) [0.0] (6,3) [0.0] (6,4) [0.0] (6,5) [0.0] (6,6) [1.0] (6,7) [0.0] (6,8) [0.0] (6,9) [0.0] (6,10) [0.0] (6,11) [0.0] (6,12) [0.0] (6,13) [0.0] (6,14) [0.0] (6,15) [0.0] (6,16) [0.0] (6,17) [0.0] (6,18) [0.0] (6,19) [0.0]

(7,0) [0.0] (7,1) [0.0] (7,2) [0.0] (7,3) [0.0] (7,4) [0.0] (7,5) [0.0] (7,6) [0.0] (7,7) [1.0] (7,8) [0.0] (7,9) [0.0] (7,10) [0.0] (7,11) [0.0] (7,12) [0.0] (7,13) [0.0] (7,14) [0.0] (7,15) [0.0] (7,16) [0.0] (7,17) [0.0] (7,18) [0.0] (7,19) [0.0]

(8,0) [0.0] (8,1) [0.0] (8,2) [0.0] (8,3) [0.0] (8,4) [0.0] (8,5) [0.0] (8,6) [0.0] (8,7) [0.0] (8,8) [1.0] (8,9) [0.0] (8,10) [0.0] (8,11) [0.0] (8,12) [0.0] (8,13) [0.0] (8,14) [0.0] (8,15) [0.0] (8,16) [0.0] (8,17) [0.0] (8,18) [0.0] (8,19) [0.0]

(9,0) [0.0] (9,1) [0.0] (9,2) [0.0] (9,3) [0.0] (9,4) [0.0] (9,5) [0.0] (9,6) [0.0] (9,7) [0.0] (9,8) [0.0] (9,9) [1.0] (9,10) [0.0] (9,11) [0.0] (9,12) [0.0] (9,13) [0.0] (9,14) [0.0] (9,15) [0.0] (9,16) [0.0] (9,17) [0.0] (9,18) [0.0] (9,19) [0.0]

(10,0) [0.0] (10,1) [0.0] (10,2) [0.0] (10,3) [0.0] (10,4) [0.0] (10,5) [0.0] (10,6) [0.0] (10,7) [0.0] (10,8) [0.0] (10,9) [0.0] (10,10) [1.0] (10,11) [0.0] (10,12) [0.0] (10,13) [0.0] (10,14) [0.0] (10,15) [0.0] (10,16) [0.0] (10,17) [0.0] (10,18) [0.0] (10,19) [0.0]

(11,0) [0.0] (11,1) [0.0] (11,2) [0.0] (11,3) [0.0] (11,4) [0.0] (11,5) [0.0] (11,6) [0.0] (11,7) [0.0] (11,8) [0.0] (11,9) [0.0] (11,10) [0.0] (11,11) [1.0] (11,12) [0.0] (11,13) [0.0] (11,14) [0.0] (11,15) [0.0] (11,16) [0.0] (11,17) [0.0] (11,18) [0.0] (11,19) [0.0]

(12,0) [0.0] (12,1) [0.0] (12,2) [0.0] (12,3) [0.0] (12,4) [0.0] (12,5) [0.0] (12,6) [0.0] (12,7) [0.0] (12,8) [0.0] (12,9) [0.0] (12,10) [0.0] (12,11) [0.0] (12,12) [1.0] (12,13) [0.0] (12,14) [0.0] (12,15) [0.0] (12,16) [0.0] (12,17) [0.0] (12,18) [0.0] (12,19) [0.0]

(13,0) [0.0] (13,1) [0.0] (13,2) [0.0] (13,3) [0.0] (13,4) [0.0] (13,5) [0.0] (13,6) [0.0] (13,7) [0.0] (13,8) [0.0] (13,9) [0.0] (13,10) [0.0] (13,11) [0.0] (13,12) [0.0] (13,13) [1.0] (13,14) [0.0] (13,15) [0.0] (13,16) [0.0] (13,17) [0.0] (13,18) [0.0] (13,19) [0.0]

(14,0) [0.0] (14,1) [0.0] (14,2) [0.0] (14,3) [0.0] (14,4) [0.0] (14,5) [0.0] (14,6) [0.0] (14,7) [0.0] (14,8) [0.0] (14,9) [0.0] (14,10) [0.0] (14,11) [0.0] (14,12) [0.0] (14,13) [0.0] (14,14) [1.0] (14,15) [0.0] (14,16) [0.0] (14,17) [0.0] (14,18) [0.0] (14,19) [0.0]

(15,0) [0.0] (15,1) [0.0] (15,2) [0.0] (15,3) [0.0] (15,4) [0.0] (15,5) [0.0] (15,6) [0.0] (15,7) [0.0] (15,8) [0.0] (15,9) [0.0] (15,10) [0.0] (15,11) [0.0] (15,12) [0.0] (15,13) [0.0] (15,14) [0.0] (15,15) [1.0] (15,16) [0.0] (15,17) [0.0] (15,18) [0.0] (15,19) [0.0]

(16,0) [0.0] (16,1) [0.0] (16,2) [0.0] (16,3) [0.0] (16,4) [0.0] (16,5) [0.0] (16,6) [0.0] (16,7) [0.0] (16,8) [0.0] (16,9) [0.0] (16,10) [0.0] (16,11) [0.0] (16,12) [0.0] (16,13) [0.0] (16,14) [0.0] (16,15) [0.0] (16,16) [1.0] (16,17) [0.0] (16,18) [0.0] (16,19) [0.0]

(17,0) [0.0] (17,1) [0.0] (17,2) [0.0] (17,3) [0.0] (17,4) [0.0] (17,5) [0.0] (17,6) [0.0] (17,7) [0.0] (17,8) [0.0] (17,9) [0.0] (17,10) [0.0] (17,11) [0.0] (17,12) [0.0] (17,13) [0.0] (17,14) [0.0] (17,15) [0.0] (17,16) [0.0] (17,17) [1.0] (17,18) [0.0] (17,19) [0.0]

(18,0) [0.0] (18,1) [0.0] (18,2) [0.0] (18,3) [0.0] (18,4) [0.0] (18,5) [0.0] (18,6) [0.0] (18,7) [0.0] (18,8) [0.0] (18,9) [0.0] (18,10) [0.0] (18,11) [0.0] (18,12) [0.0] (18,13) [0.0] (18,14) [0.0] (18,15) [0.0] (18,16) [0.0] (18,17) [0.0] (18,18) [1.0] (18,19) [0.0]

(19,0) [0.0] (19,1) [0.0] (19,2) [0.0] (19,3) [0.0] (19,4) [0.0] (19,5) [0.0] (19,6) [0.0] (19,7) [0.0] (19,8) [0.0] (19,9) [0.0] (19,10) [0.0] (19,11) [0.0] (19,12) [0.0] (19,13) [0.0] (19,14) [0.0] (19,15) [0.0] (19,16) [0.0] (19,17) [0.0] (19,18) [0.0] (19,19) [1.0]

};
\end{axis}
\end{tikzpicture}
    \centering
    \caption{242 sub-carriers, accuracy: 98.66\%}
    \end{subfigure}
    \hfill
    \begin{subfigure}[t]{0.49\columnwidth}
    \centering
    \setlength\fwidth{.7\columnwidth}
    \setlength\fheight{0.6\columnwidth}
    \begin{tikzpicture}
\pgfplotsset{every tick label/.append style={font=\tiny}}

\begin{axis}[
enlargelimits=false,
colorbar,
colormap/Purples,
width=\fwidth,
height=\fheight,
at={(0\fwidth,0\fheight)},
scale only axis,
tick align=inside,
xlabel={Predicted Activity},
xmin=-0.5,
xmax=19.5,
xtick style={draw=none},
xlabel style={font=\scriptsize\color{white!15!black}},
ylabel style={font=\scriptsize\color{white!15!black}},
ylabel={Actual Activity},
ymin=-0.5,
ymax=19.5,
xlabel shift=-5pt,
ylabel shift=-5pt,
ytick style={draw=none},
axis background/.style={fill=white},
colorbar horizontal,
colorbar style={
at={(0,1.05)},               
anchor=below south west,    
width=\pgfkeysvalueof{/pgfplots/parent axis width},
xtick={0, 0.5, 1},
xmin=0,
xmax=1,
axis x line*=top,
xticklabel shift=-1pt,
point meta min=0,
point meta max=1,
},
colorbar/width=2mm,
]
\addplot [matrix plot,point meta=explicit]
 coordinates {
(0,0) [0.9] (0,1) [0.1] (0,2) [0.0] (0,3) [0.0] (0,4) [0.0] (0,5) [0.0] (0,6) [0.0] (0,7) [0.0] (0,8) [0.0] (0,9) [0.0] (0,10) [0.0] (0,11) [0.0] (0,12) [0.0] (0,13) [0.0] (0,14) [0.0] (0,15) [0.0] (0,16) [0.0] (0,17) [0.0] (0,18) [0.0] (0,19) [0.0]

(1,0) [0.0] (1,1) [0.9] (1,2) [0.0] (1,3) [0.0] (1,4) [0.0] (1,5) [0.0] (1,6) [0.0] (1,7) [0.0] (1,8) [0.0] (1,9) [0.0] (1,10) [0.0] (1,11) [0.0] (1,12) [0.0] (1,13) [0.0] (1,14) [0.0] (1,15) [0.0] (1,16) [0.0] (1,17) [0.0] (1,18) [0.0] (1,19) [0.0]

(2,0) [0.1] (2,1) [0.0] (2,2) [0.7] (2,3) [0.0] (2,4) [0.0] (2,5) [0.3] (2,6) [0.0] (2,7) [0.0] (2,8) [0.0] (2,9) [0.0] (2,10) [0.0] (2,11) [0.0] (2,12) [0.0] (2,13) [0.0] (2,14) [0.0] (2,15) [0.0] (2,16) [0.0] (2,17) [0.0] (2,18) [0.0] (2,19) [0.0]

(3,0) [0.0] (3,1) [0.0] (3,2) [0.0] (3,3) [0.9] (3,4) [0.0] (3,5) [0.0] (3,6) [0.0] (3,7) [0.0] (3,8) [0.0] (3,9) [0.0] (3,10) [0.0] (3,11) [0.0] (3,12) [0.0] (3,13) [0.0] (3,14) [0.0] (3,15) [0.0] (3,16) [0.0] (3,17) [0.0] (3,18) [0.0] (3,19) [0.0]

(4,0) [0.0] (4,1) [0.0] (4,2) [0.0] (4,3) [0.0] (4,4) [0.9] (4,5) [0.0] (4,6) [0.0] (4,7) [0.0] (4,8) [0.0] (4,9) [0.0] (4,10) [0.0] (4,11) [0.0] (4,12) [0.0] (4,13) [0.0] (4,14) [0.0] (4,15) [0.0] (4,16) [0.0] (4,17) [0.0] (4,18) [0.0] (4,19) [0.0]

(5,0) [0.0] (5,1) [0.0] (5,2) [0.2] (5,3) [0.0] (5,4) [0.0] (5,5) [0.6] (5,6) [0.0] (5,7) [0.0] (5,8) [0.0] (5,9) [0.0] (5,10) [0.0] (5,11) [0.0] (5,12) [0.0] (5,13) [0.0] (5,14) [0.0] (5,15) [0.0] (5,16) [0.0] (5,17) [0.0] (5,18) [0.0] (5,19) [0.0]

(6,0) [0.0] (6,1) [0.0] (6,2) [0.0] (6,3) [0.0] (6,4) [0.0] (6,5) [0.0] (6,6) [0.5] (6,7) [0.0] (6,8) [0.0] (6,9) [0.1] (6,10) [0.0] (6,11) [0.2] (6,12) [0.0] (6,13) [0.0] (6,14) [0.0] (6,15) [0.0] (6,16) [0.0] (6,17) [0.0] (6,18) [0.1] (6,19) [0.1]

(7,0) [0.0] (7,1) [0.0] (7,2) [0.0] (7,3) [0.0] (7,4) [0.0] (7,5) [0.0] (7,6) [0.0] (7,7) [1.0] (7,8) [0.0] (7,9) [0.0] (7,10) [0.0] (7,11) [0.0] (7,12) [0.0] (7,13) [0.0] (7,14) [0.0] (7,15) [0.0] (7,16) [0.0] (7,17) [0.0] (7,18) [0.0] (7,19) [0.0]

(8,0) [0.0] (8,1) [0.0] (8,2) [0.0] (8,3) [0.0] (8,4) [0.0] (8,5) [0.0] (8,6) [0.0] (8,7) [0.0] (8,8) [1.0] (8,9) [0.0] (8,10) [0.0] (8,11) [0.0] (8,12) [0.0] (8,13) [0.0] (8,14) [0.0] (8,15) [0.0] (8,16) [0.0] (8,17) [0.0] (8,18) [0.0] (8,19) [0.0]

(9,0) [0.0] (9,1) [0.0] (9,2) [0.0] (9,3) [0.0] (9,4) [0.0] (9,5) [0.0] (9,6) [0.1] (9,7) [0.0] (9,8) [0.0] (9,9) [0.3] (9,10) [0.2] (9,11) [0.0] (9,12) [0.1] (9,13) [0.0] (9,14) [0.1] (9,15) [0.1] (9,16) [0.0] (9,17) [0.4] (9,18) [0.1] (9,19) [0.1]

(10,0) [0.0] (10,1) [0.0] (10,2) [0.0] (10,3) [0.0] (10,4) [0.0] (10,5) [0.0] (10,6) [0.0] (10,7) [0.0] (10,8) [0.0] (10,9) [0.1] (10,10) [0.0] (10,11) [0.0] (10,12) [0.1] (10,13) [0.0] (10,14) [0.2] (10,15) [0.2] (10,16) [0.1] (10,17) [0.2] (10,18) [0.0] (10,19) [0.0]

(11,0) [0.0] (11,1) [0.0] (11,2) [0.0] (11,3) [0.0] (11,4) [0.0] (11,5) [0.0] (11,6) [0.3] (11,7) [0.0] (11,8) [0.0] (11,9) [0.1] (11,10) [0.0] (11,11) [0.6] (11,12) [0.0] (11,13) [0.0] (11,14) [0.0] (11,15) [0.0] (11,16) [0.0] (11,17) [0.0] (11,18) [0.1] (11,19) [0.1]

(12,0) [0.0] (12,1) [0.0] (12,2) [0.0] (12,3) [0.0] (12,4) [0.0] (12,5) [0.0] (12,6) [0.0] (12,7) [0.0] (12,8) [0.0] (12,9) [0.1] (12,10) [0.2] (12,11) [0.0] (12,12) [0.5] (12,13) [0.0] (12,14) [0.1] (12,15) [0.0] (12,16) [0.0] (12,17) [0.0] (12,18) [0.1] (12,19) [0.0]

(13,0) [0.0] (13,1) [0.0] (13,2) [0.0] (13,3) [0.0] (13,4) [0.0] (13,5) [0.0] (13,6) [0.0] (13,7) [0.0] (13,8) [0.0] (13,9) [0.0] (13,10) [0.2] (13,11) [0.0] (13,12) [0.0] (13,13) [0.5] (13,14) [0.1] (13,15) [0.2] (13,16) [0.3] (13,17) [0.0] (13,18) [0.0] (13,19) [0.0]

(14,0) [0.0] (14,1) [0.0] (14,2) [0.0] (14,3) [0.0] (14,4) [0.0] (14,5) [0.0] (14,6) [0.0] (14,7) [0.0] (14,8) [0.0] (14,9) [0.1] (14,10) [0.2] (14,11) [0.0] (14,12) [0.0] (14,13) [0.1] (14,14) [0.2] (14,15) [0.2] (14,16) [0.0] (14,17) [0.1] (14,18) [0.0] (14,19) [0.0]

(15,0) [0.0] (15,1) [0.0] (15,2) [0.0] (15,3) [0.0] (15,4) [0.0] (15,5) [0.0] (15,6) [0.0] (15,7) [0.0] (15,8) [0.0] (15,9) [0.1] (15,10) [0.1] (15,11) [0.0] (15,12) [0.1] (15,13) [0.1] (15,14) [0.1] (15,15) [0.1] (15,16) [0.1] (15,17) [0.1] (15,18) [0.0] (15,19) [0.0]

(16,0) [0.0] (16,1) [0.0] (16,2) [0.0] (16,3) [0.0] (16,4) [0.0] (16,5) [0.0] (16,6) [0.0] (16,7) [0.0] (16,8) [0.0] (16,9) [0.0] (16,10) [0.1] (16,11) [0.0] (16,12) [0.0] (16,13) [0.2] (16,14) [0.1] (16,15) [0.1] (16,16) [0.4] (16,17) [0.1] (16,18) [0.0] (16,19) [0.0]

(17,0) [0.0] (17,1) [0.0] (17,2) [0.0] (17,3) [0.0] (17,4) [0.0] (17,5) [0.0] (17,6) [0.0] (17,7) [0.0] (17,8) [0.0] (17,9) [0.0] (17,10) [0.0] (17,11) [0.0] (17,12) [0.0] (17,13) [0.1] (17,14) [0.0] (17,15) [0.0] (17,16) [0.0] (17,17) [0.0] (17,18) [0.0] (17,19) [0.0]

(18,0) [0.0] (18,1) [0.0] (18,2) [0.0] (18,3) [0.0] (18,4) [0.0] (18,5) [0.0] (18,6) [0.0] (18,7) [0.0] (18,8) [0.0] (18,9) [0.1] (18,10) [0.1] (18,11) [0.0] (18,12) [0.2] (18,13) [0.0] (18,14) [0.0] (18,15) [0.0] (18,16) [0.0] (18,17) [0.1] (18,18) [0.4] (18,19) [0.1]

(19,0) [0.0] (19,1) [0.0] (19,2) [0.0] (19,3) [0.0] (19,4) [0.0] (19,5) [0.0] (19,6) [0.1] (19,7) [0.0] (19,8) [0.0] (19,9) [0.0] (19,10) [0.0] (19,11) [0.2] (19,12) [0.0] (19,13) [0.0] (19,14) [0.0] (19,15) [0.0] (19,16) [0.0] (19,17) [0.0] (19,18) [0.3] (19,19) [0.6]

};
\end{axis}
\end{tikzpicture}
    \centering
    \caption{20 sub-carriers, accuracy: 64.37\%}
    \label{fig:cm_csi}
    \end{subfigure}
     
    \setlength\abovecaptionskip{0.2cm}
    \caption{Confusion matrices of baseline \gls{cnn} with 242 and 20 subcarriers respectively (in classroom, with monitor M2).} 
    \label{fig:cnn_subcarrier_242_vs_20}
\end{figure}
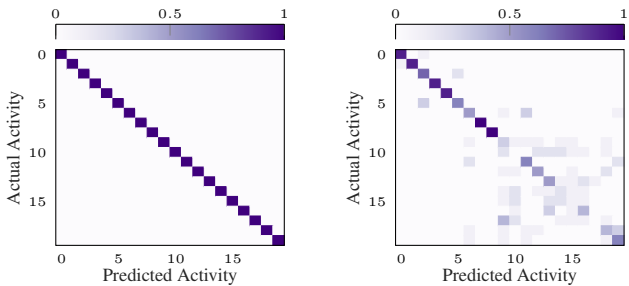
Figure \ref{fig:subcarrier_resolution} shows the performance of \FW as a function of the number of subcarriers. The first consecutive 20, 40, 80 and 160 and 242 subcarriers are considered to emulate a sensing system with a lower bandwidth - thus, less number of consecutive subcarriers. The results show that the average performance of monitors decrease to 84.92\% when the number of subcarriers decrease from 242 to 80, and it goes down to 62.76\% with a percent decrease of 34.94 when we switch to only 20 subcarriers. Figure \ref{fig:cnn_subcarrier_242_vs_20} presents the confusion matrices of baseline \gls{cnn} when trained with 242 and 20 subcarriers respectively where they achieved an accuracy of 98.66\% and 64.37\% respectively. It is evident that, with only 20 subcarriers, the model gets confused with few activities whereas it performs comparatively better in other activities. The top three classes which are hardest to distinguish at lower subcarrier resolutions are: wash hands (index 17), rotate (index 10), and brush (index 15). However, it is noticeable that when we switch to 20 subcarriers from 242, the input tensor dimension reduces by 12 times from $50\times 242\times 2= 24200$ to $50\times 20\times 2= 2000$ and still achieve around 64\% accuracy on an average in the considered scenarios. \smallskip

\begin{figure}[ht]
	\centering
	\includegraphics[width=.93\columnwidth]{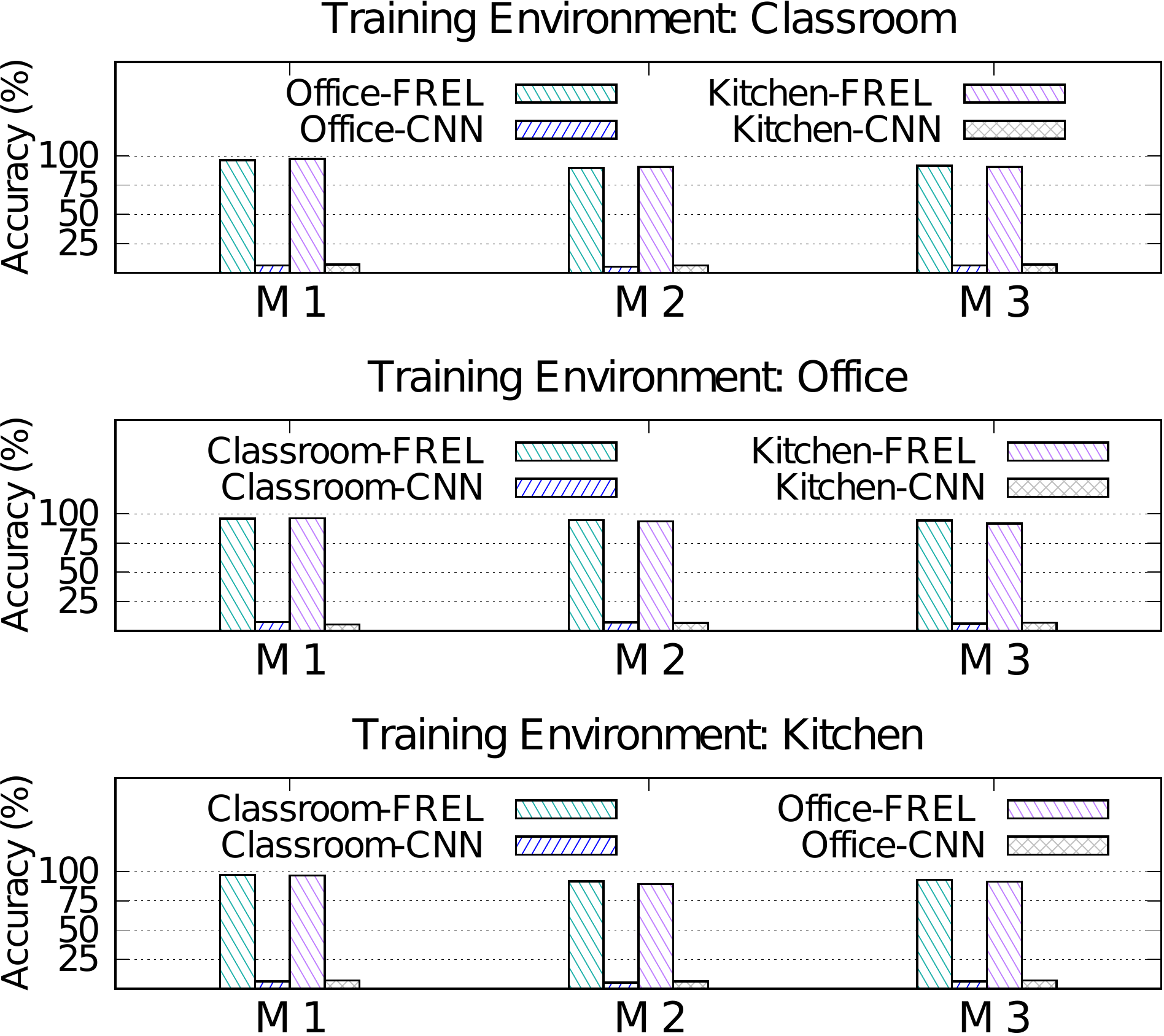}
	\caption{Performance of \FR in simultaneous activity sensing with new untrained environments.  }
	\label{fig:FREL_vs_CNN}
\end{figure}

\begin{figure}[ht]
	\centering
	\includegraphics[width=.93\columnwidth]{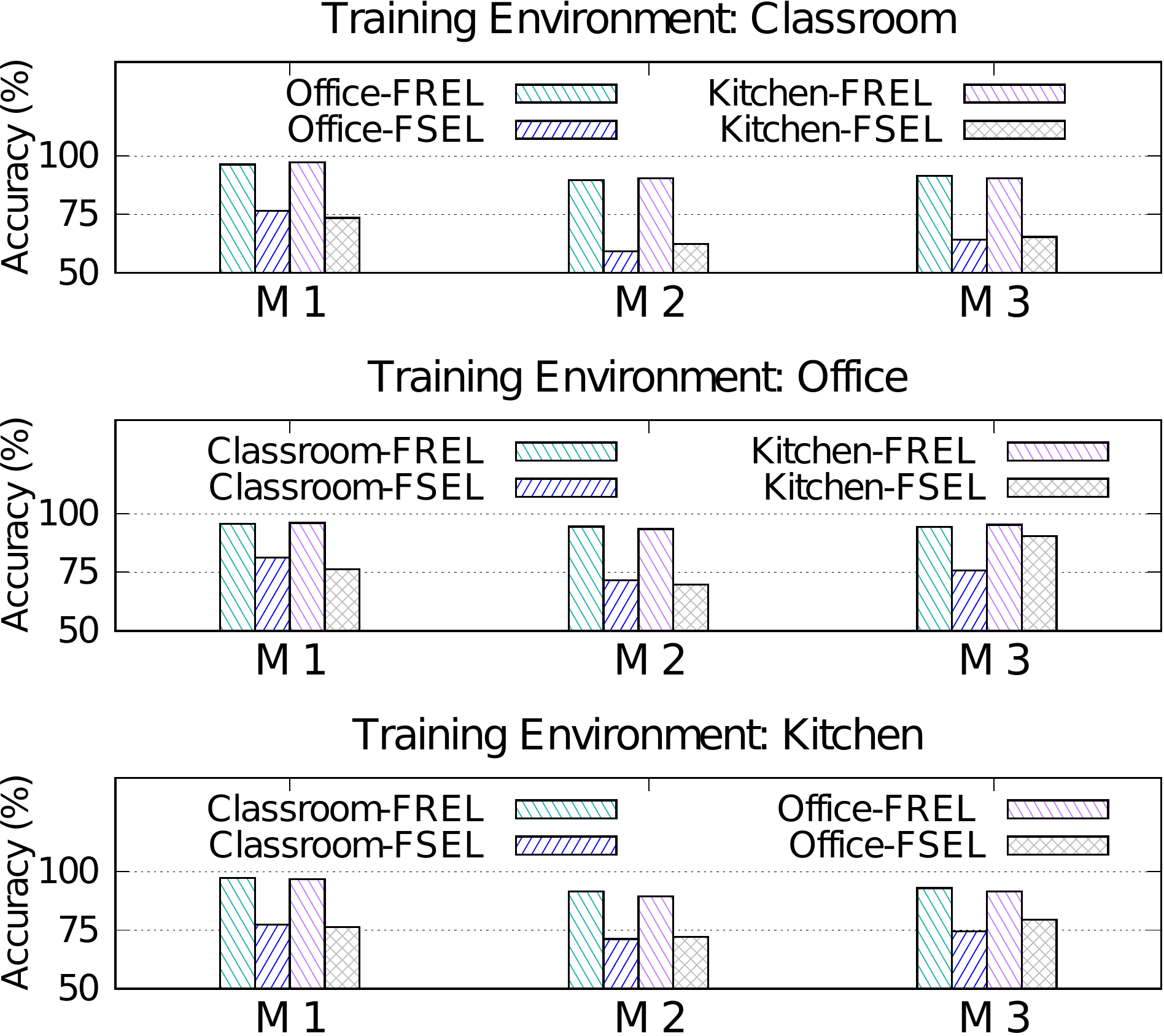}
	\caption{Performance comparison of \FR with \gls{fsel} in new untrained environments. }
	\label{fig:FREL_vs_FSE}
\end{figure}

\textbf{\FW performance with \FR:} Even though the performance of the traditional \gls{cnn} is quite good, they fail to generalize the environments or subjects. In such instances for generalizing to new environments and subjects, \FR excels in comparison to the traditional \gls{cnn}. The performance of the \FR in new untrained environment are presented and compared with traditional \gls{cnn} and \gls{fsel}\cite{tian2020rethinking} in Figure \ref{fig:FREL_vs_CNN} and Figure \ref{fig:FREL_vs_FSE} respectively. The results show that the overall performance of \FW with \FR in new untrained environments improves by 86.06\%, 87.56\% and  86.90\%  when the embedding network is trained in classroom, office and kitchen respectively in comparison to the baseline \gls{cnn}. The highest performance achieved by \FR in any new untrained environment is 97.24\% in the kitchen with monitor M1 whereas the lowest accuracy is 89.33\% in the kitchen with monitor M2. Thus, it demonstrates the robustness, reliability and strong adaptive capability of \FR to new environments. As shown in Figure \ref{fig:FREL_vs_FSE},  \FR surpasses the \gls{fsel} by 25.71\%,  17.43\%, and 15.56\% respectively when the \FR is trained in the classroom, office and kitchen and tested on other corresponding environments.  

\vspace{-0.2cm}

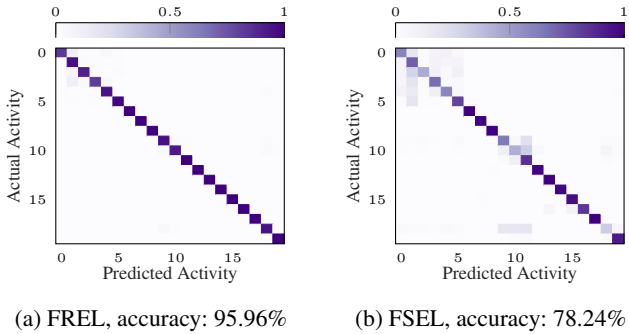
\begin{figure}[h]
    \centering
    \begin{subfigure}[t]{0.49\columnwidth}
    \centering
    \setlength\fwidth{.7\columnwidth}
    \setlength\fheight{0.6\columnwidth}
    \begin{tikzpicture}
\pgfplotsset{every tick label/.append style={font=\tiny}}

\begin{axis}[
enlargelimits=false,
colorbar,
colormap/Purples,
width=\fwidth,
height=\fheight,
at={(0\fwidth,0\fheight)},
scale only axis,
tick align=inside,
xlabel={Predicted Activity},
xmin=-0.5,
xmax=19.5,
xtick style={draw=none},
xlabel style={font=\scriptsize\color{white!15!black}},
ylabel style={font=\scriptsize\color{white!15!black}},
ylabel={Actual Activity},
ymin=-0.5,
ymax=19.5,
xlabel shift=-5pt,
ylabel shift=-5pt,
ytick style={draw=none},
axis background/.style={fill=white},
colorbar horizontal,
colorbar style={
at={(0,1.05)},               
anchor=below south west,    
width=\pgfkeysvalueof{/pgfplots/parent axis width},
xtick={0, 0.5, 1},
xmin=0,
xmax=1,
axis x line*=top,
xticklabel shift=-1pt,
point meta min=0,
point meta max=1,
},
colorbar/width=2mm,
]
\addplot [matrix plot,point meta=explicit]
 coordinates {
(0,0) [0.82] (0,1) [0.01] (0,2) [0.01] (0,3) [0.0] (0,4) [0.01] (0,5) [0.0] (0,6) [0.0] (0,7) [0.0] (0,8) [0.0] (0,9) [0.0] (0,10) [0.0] (0,11) [0.0] (0,12) [0.0] (0,13) [0.0] (0,14) [0.0] (0,15) [0.0] (0,16) [0.0] (0,17) [0.0] (0,18) [0.0] (0,19) [0.0]

(1,0) [0.11] (1,1) [0.95] (1,2) [0.06] (1,3) [0.11] (1,4) [0.0] (1,5) [0.01] (1,6) [0.0] (1,7) [0.0] (1,8) [0.0] (1,9) [0.0] (1,10) [0.0] (1,11) [0.0] (1,12) [0.0] (1,13) [0.0] (1,14) [0.0] (1,15) [0.0] (1,16) [0.0] (1,17) [0.0] (1,18) [0.0] (1,19) [0.0]

(2,0) [0.01] (2,1) [0.02] (2,2) [0.9] (2,3) [0.03] (2,4) [0.0] (2,5) [0.0] (2,6) [0.0] (2,7) [0.0] (2,8) [0.0] (2,9) [0.0] (2,10) [0.0] (2,11) [0.0] (2,12) [0.0] (2,13) [0.0] (2,14) [0.0] (2,15) [0.0] (2,16) [0.0] (2,17) [0.0] (2,18) [0.0] (2,19) [0.0]

(3,0) [0.01] (3,1) [0.01] (3,2) [0.01] (3,3) [0.81] (3,4) [0.01] (3,5) [0.01] (3,6) [0.0] (3,7) [0.0] (3,8) [0.0] (3,9) [0.0] (3,10) [0.0] (3,11) [0.0] (3,12) [0.0] (3,13) [0.0] (3,14) [0.0] (3,15) [0.0] (3,16) [0.0] (3,17) [0.0] (3,18) [0.0] (3,19) [0.0]

(4,0) [0.04] (4,1) [0.01] (4,2) [0.0] (4,3) [0.04] (4,4) [0.94] (4,5) [0.0] (4,6) [0.0] (4,7) [0.0] (4,8) [0.0] (4,9) [0.0] (4,10) [0.0] (4,11) [0.0] (4,12) [0.0] (4,13) [0.0] (4,14) [0.0] (4,15) [0.0] (4,16) [0.0] (4,17) [0.0] (4,18) [0.0] (4,19) [0.0]

(5,0) [0.01] (5,1) [0.01] (5,2) [0.02] (5,3) [0.0] (5,4) [0.02] (5,5) [0.98] (5,6) [0.0] (5,7) [0.0] (5,8) [0.0] (5,9) [0.0] (5,10) [0.0] (5,11) [0.0] (5,12) [0.0] (5,13) [0.0] (5,14) [0.0] (5,15) [0.0] (5,16) [0.0] (5,17) [0.0] (5,18) [0.0] (5,19) [0.0]

(6,0) [0.0] (6,1) [0.0] (6,2) [0.0] (6,3) [0.0] (6,4) [0.0] (6,5) [0.0] (6,6) [1.0] (6,7) [0.0] (6,8) [0.0] (6,9) [0.0] (6,10) [0.0] (6,11) [0.0] (6,12) [0.0] (6,13) [0.0] (6,14) [0.0] (6,15) [0.0] (6,16) [0.0] (6,17) [0.0] (6,18) [0.0] (6,19) [0.0]

(7,0) [0.0] (7,1) [0.0] (7,2) [0.0] (7,3) [0.0] (7,4) [0.0] (7,5) [0.0] (7,6) [0.0] (7,7) [1.0] (7,8) [0.0] (7,9) [0.0] (7,10) [0.0] (7,11) [0.0] (7,12) [0.0] (7,13) [0.0] (7,14) [0.0] (7,15) [0.0] (7,16) [0.0] (7,17) [0.0] (7,18) [0.0] (7,19) [0.0]

(8,0) [0.0] (8,1) [0.0] (8,2) [0.0] (8,3) [0.0] (8,4) [0.0] (8,5) [0.0] (8,6) [0.0] (8,7) [0.0] (8,8) [1.0] (8,9) [0.0] (8,10) [0.0] (8,11) [0.0] (8,12) [0.0] (8,13) [0.0] (8,14) [0.0] (8,15) [0.0] (8,16) [0.0] (8,17) [0.0] (8,18) [0.0] (8,19) [0.0]

(9,0) [0.0] (9,1) [0.0] (9,2) [0.0] (9,3) [0.0] (9,4) [0.0] (9,5) [0.0] (9,6) [0.0] (9,7) [0.0] (9,8) [0.0] (9,9) [0.95] (9,10) [0.03] (9,11) [0.0] (9,12) [0.0] (9,13) [0.0] (9,14) [0.0] (9,15) [0.0] (9,16) [0.0] (9,17) [0.0] (9,18) [0.04] (9,19) [0.0]

(10,0) [0.0] (10,1) [0.0] (10,2) [0.0] (10,3) [0.0] (10,4) [0.0] (10,5) [0.0] (10,6) [0.0] (10,7) [0.0] (10,8) [0.0] (10,9) [0.02] (10,10) [0.92] (10,11) [0.0] (10,12) [0.0] (10,13) [0.0] (10,14) [0.0] (10,15) [0.0] (10,16) [0.0] (10,17) [0.0] (10,18) [0.01] (10,19) [0.0]

(11,0) [0.0] (11,1) [0.0] (11,2) [0.0] (11,3) [0.0] (11,4) [0.0] (11,5) [0.0] (11,6) [0.0] (11,7) [0.0] (11,8) [0.0] (11,9) [0.01] (11,10) [0.01] (11,11) [1.0] (11,12) [0.0] (11,13) [0.0] (11,14) [0.0] (11,15) [0.0] (11,16) [0.0] (11,17) [0.0] (11,18) [0.0] (11,19) [0.0]

(12,0) [0.0] (12,1) [0.0] (12,2) [0.0] (12,3) [0.0] (12,4) [0.0] (12,5) [0.0] (12,6) [0.0] (12,7) [0.0] (12,8) [0.0] (12,9) [0.0] (12,10) [0.0] (12,11) [0.0] (12,12) [1.0] (12,13) [0.0] (12,14) [0.0] (12,15) [0.0] (12,16) [0.0] (12,17) [0.0] (12,18) [0.0] (12,19) [0.0]

(13,0) [0.0] (13,1) [0.0] (13,2) [0.0] (13,3) [0.0] (13,4) [0.0] (13,5) [0.0] (13,6) [0.0] (13,7) [0.0] (13,8) [0.0] (13,9) [0.0] (13,10) [0.0] (13,11) [0.0] (13,12) [0.0] (13,13) [1.0] (13,14) [0.0] (13,15) [0.0] (13,16) [0.0] (13,17) [0.0] (13,18) [0.0] (13,19) [0.0]

(14,0) [0.0] (14,1) [0.0] (14,2) [0.0] (14,3) [0.0] (14,4) [0.0] (14,5) [0.0] (14,6) [0.0] (14,7) [0.0] (14,8) [0.0] (14,9) [0.0] (14,10) [0.0] (14,11) [0.0] (14,12) [0.0] (14,13) [0.0] (14,14) [1.0] (14,15) [0.0] (14,16) [0.0] (14,17) [0.0] (14,18) [0.0] (14,19) [0.0]

(15,0) [0.0] (15,1) [0.0] (15,2) [0.0] (15,3) [0.0] (15,4) [0.0] (15,5) [0.0] (15,6) [0.0] (15,7) [0.0] (15,8) [0.0] (15,9) [0.0] (15,10) [0.0] (15,11) [0.0] (15,12) [0.0] (15,13) [0.0] (15,14) [0.0] (15,15) [1.0] (15,16) [0.0] (15,17) [0.0] (15,18) [0.0] (15,19) [0.0]

(16,0) [0.0] (16,1) [0.0] (16,2) [0.0] (16,3) [0.0] (16,4) [0.0] (16,5) [0.0] (16,6) [0.0] (16,7) [0.0] (16,8) [0.0] (16,9) [0.0] (16,10) [0.0] (16,11) [0.0] (16,12) [0.0] (16,13) [0.0] (16,14) [0.0] (16,15) [0.0] (16,16) [1.0] (16,17) [0.0] (16,18) [0.0] (16,19) [0.0]

(17,0) [0.0] (17,1) [0.0] (17,2) [0.0] (17,3) [0.0] (17,4) [0.0] (17,5) [0.0] (17,6) [0.0] (17,7) [0.0] (17,8) [0.0] (17,9) [0.0] (17,10) [0.0] (17,11) [0.0] (17,12) [0.0] (17,13) [0.0] (17,14) [0.0] (17,15) [0.0] (17,16) [0.0] (17,17) [1.0] (17,18) [0.0] (17,19) [0.0]

(18,0) [0.0] (18,1) [0.0] (18,2) [0.0] (18,3) [0.0] (18,4) [0.02] (18,5) [0.0] (18,6) [0.0] (18,7) [0.0] (18,8) [0.0] (18,9) [0.02] (18,10) [0.03] (18,11) [0.0] (18,12) [0.0] (18,13) [0.0] (18,14) [0.0] (18,15) [0.0] (18,16) [0.0] (18,17) [0.0] (18,18) [0.94] (18,19) [0.0]

(19,0) [0.0] (19,1) [0.0] (19,2) [0.0] (19,3) [0.0] (19,4) [0.0] (19,5) [0.0] (19,6) [0.0] (19,7) [0.0] (19,8) [0.0] (19,9) [0.0] (19,10) [0.01] (19,11) [0.0] (19,12) [0.0] (19,13) [0.0] (19,14) [0.0] (19,15) [0.0] (19,16) [0.0] (19,17) [0.0] (19,18) [0.0] (19,19) [1.0]

};
\end{axis}
\end{tikzpicture}
    \centering
    \caption{\FR, accuracy: 95.96\%}
    \end{subfigure}
    \hfill
    \begin{subfigure}[t]{0.49\columnwidth}
    \centering
    \setlength\fwidth{.7\columnwidth}
    \setlength\fheight{0.6\columnwidth}
    \begin{tikzpicture}
\pgfplotsset{every tick label/.append style={font=\tiny}}

\begin{axis}[
enlargelimits=false,
colorbar,
colormap/Purples,
width=\fwidth,
height=\fheight,
at={(0\fwidth,0\fheight)},
scale only axis,
tick align=inside,
xlabel={Predicted Activity},
xmin=-0.5,
xmax=19.5,
xtick style={draw=none},
xlabel style={font=\scriptsize\color{white!15!black}},
ylabel style={font=\scriptsize\color{white!15!black}},
ylabel={Actual Activity},
ymin=-0.5,
ymax=19.5,
xlabel shift=-5pt,
ylabel shift=-5pt,
ytick style={draw=none},
axis background/.style={fill=white},
colorbar horizontal,
colorbar style={
at={(0,1.05)},               
anchor=below south west,    
width=\pgfkeysvalueof{/pgfplots/parent axis width},
xtick={0, 0.5, 1},
xmin=0,
xmax=1,
axis x line*=top,
xticklabel shift=-1pt,
point meta min=0,
point meta max=1,
},
colorbar/width=2mm,
]
\addplot [matrix plot,point meta=explicit]
 coordinates {
(0,0) [0.59] (0,1) [0.07] (0,2) [0.06] (0,3) [0.05] (0,4) [0.09] (0,5) [0.01] (0,6) [0.0] (0,7) [0.0] (0,8) [0.0] (0,9) [0.01] (0,10) [0.01] (0,11) [0.0] (0,12) [0.0] (0,13) [0.0] (0,14) [0.0] (0,15) [0.0] (0,16) [0.0] (0,17) [0.0] (0,18) [0.03] (0,19) [0.0]

(1,0) [0.17] (1,1) [0.71] (1,2) [0.27] (1,3) [0.19] (1,4) [0.14] (1,5) [0.18] (1,6) [0.0] (1,7) [0.0] (1,8) [0.0] (1,9) [0.01] (1,10) [0.0] (1,11) [0.0] (1,12) [0.0] (1,13) [0.0] (1,14) [0.0] (1,15) [0.0] (1,16) [0.0] (1,17) [0.0] (1,18) [0.02] (1,19) [0.0]

(2,0) [0.02] (2,1) [0.01] (2,2) [0.44] (2,3) [0.02] (2,4) [0.01] (2,5) [0.01] (2,6) [0.0] (2,7) [0.0] (2,8) [0.0] (2,9) [0.02] (2,10) [0.01] (2,11) [0.0] (2,12) [0.0] (2,13) [0.0] (2,14) [0.0] (2,15) [0.0] (2,16) [0.0] (2,17) [0.0] (2,18) [0.02] (2,19) [0.0]

(3,0) [0.07] (3,1) [0.08] (3,2) [0.09] (3,3) [0.67] (3,4) [0.11] (3,5) [0.01] (3,6) [0.0] (3,7) [0.0] (3,8) [0.0] (3,9) [0.01] (3,10) [0.0] (3,11) [0.0] (3,12) [0.0] (3,13) [0.0] (3,14) [0.0] (3,15) [0.0] (3,16) [0.0] (3,17) [0.0] (3,18) [0.01] (3,19) [0.0]

(4,0) [0.09] (4,1) [0.06] (4,2) [0.03] (4,3) [0.03] (4,4) [0.59] (4,5) [0.0] (4,6) [0.0] (4,7) [0.0] (4,8) [0.0] (4,9) [0.0] (4,10) [0.0] (4,11) [0.0] (4,12) [0.0] (4,13) [0.0] (4,14) [0.0] (4,15) [0.0] (4,16) [0.0] (4,17) [0.0] (4,18) [0.03] (4,19) [0.0]

(5,0) [0.04] (5,1) [0.07] (5,2) [0.1] (5,3) [0.03] (5,4) [0.04] (5,5) [0.79] (5,6) [0.0] (5,7) [0.0] (5,8) [0.0] (5,9) [0.03] (5,10) [0.01] (5,11) [0.0] (5,12) [0.0] (5,13) [0.0] (5,14) [0.0] (5,15) [0.0] (5,16) [0.0] (5,17) [0.0] (5,18) [0.02] (5,19) [0.01]

(6,0) [0.0] (6,1) [0.0] (6,2) [0.0] (6,3) [0.0] (6,4) [0.0] (6,5) [0.0] (6,6) [1.0] (6,7) [0.01] (6,8) [0.0] (6,9) [0.0] (6,10) [0.0] (6,11) [0.0] (6,12) [0.0] (6,13) [0.0] (6,14) [0.0] (6,15) [0.0] (6,16) [0.0] (6,17) [0.0] (6,18) [0.0] (6,19) [0.0]

(7,0) [0.0] (7,1) [0.0] (7,2) [0.0] (7,3) [0.0] (7,4) [0.0] (7,5) [0.0] (7,6) [0.0] (7,7) [0.99] (7,8) [0.0] (7,9) [0.0] (7,10) [0.0] (7,11) [0.0] (7,12) [0.0] (7,13) [0.0] (7,14) [0.0] (7,15) [0.01] (7,16) [0.0] (7,17) [0.0] (7,18) [0.0] (7,19) [0.0]

(8,0) [0.0] (8,1) [0.0] (8,2) [0.0] (8,3) [0.0] (8,4) [0.0] (8,5) [0.0] (8,6) [0.0] (8,7) [0.0] (8,8) [1.0] (8,9) [0.0] (8,10) [0.0] (8,11) [0.0] (8,12) [0.01] (8,13) [0.0] (8,14) [0.0] (8,15) [0.0] (8,16) [0.0] (8,17) [0.0] (8,18) [0.0] (8,19) [0.0]

(9,0) [0.0] (9,1) [0.0] (9,2) [0.0] (9,3) [0.0] (9,4) [0.0] (9,5) [0.0] (9,6) [0.0] (9,7) [0.0] (9,8) [0.0] (9,9) [0.63] (9,10) [0.11] (9,11) [0.02] (9,12) [0.0] (9,13) [0.0] (9,14) [0.0] (9,15) [0.0] (9,16) [0.0] (9,17) [0.0] (9,18) [0.17] (9,19) [0.0]

(10,0) [0.0] (10,1) [0.0] (10,2) [0.0] (10,3) [0.0] (10,4) [0.0] (10,5) [0.0] (10,6) [0.0] (10,7) [0.0] (10,8) [0.0] (10,9) [0.04] (10,10) [0.43] (10,11) [0.11] (10,12) [0.0] (10,13) [0.0] (10,14) [0.0] (10,15) [0.0] (10,16) [0.0] (10,17) [0.0] (10,18) [0.17] (10,19) [0.02]

(11,0) [0.0] (11,1) [0.0] (11,2) [0.0] (11,3) [0.0] (11,4) [0.0] (11,5) [0.0] (11,6) [0.0] (11,7) [0.0] (11,8) [0.0] (11,9) [0.21] (11,10) [0.3] (11,11) [0.85] (11,12) [0.0] (11,13) [0.0] (11,14) [0.0] (11,15) [0.0] (11,16) [0.0] (11,17) [0.0] (11,18) [0.18] (11,19) [0.03]

(12,0) [0.0] (12,1) [0.0] (12,2) [0.0] (12,3) [0.0] (12,4) [0.0] (12,5) [0.0] (12,6) [0.0] (12,7) [0.0] (12,8) [0.0] (12,9) [0.0] (12,10) [0.04] (12,11) [0.0] (12,12) [0.99] (12,13) [0.0] (12,14) [0.0] (12,15) [0.0] (12,16) [0.0] (12,17) [0.0] (12,18) [0.0] (12,19) [0.0]

(13,0) [0.0] (13,1) [0.0] (13,2) [0.0] (13,3) [0.0] (13,4) [0.0] (13,5) [0.0] (13,6) [0.0] (13,7) [0.0] (13,8) [0.0] (13,9) [0.0] (13,10) [0.0] (13,11) [0.0] (13,12) [0.0] (13,13) [1.0] (13,14) [0.01] (13,15) [0.0] (13,16) [0.05] (13,17) [0.0] (13,18) [0.0] (13,19) [0.0]

(14,0) [0.0] (14,1) [0.0] (14,2) [0.0] (14,3) [0.0] (14,4) [0.0] (14,5) [0.0] (14,6) [0.0] (14,7) [0.0] (14,8) [0.0] (14,9) [0.0] (14,10) [0.0] (14,11) [0.0] (14,12) [0.0] (14,13) [0.0] (14,14) [0.98] (14,15) [0.02] (14,16) [0.01] (14,17) [0.0] (14,18) [0.0] (14,19) [0.0]

(15,0) [0.0] (15,1) [0.0] (15,2) [0.0] (15,3) [0.0] (15,4) [0.0] (15,5) [0.0] (15,6) [0.0] (15,7) [0.0] (15,8) [0.0] (15,9) [0.0] (15,10) [0.0] (15,11) [0.0] (15,12) [0.0] (15,13) [0.0] (15,14) [0.0] (15,15) [0.96] (15,16) [0.09] (15,17) [0.0] (15,18) [0.0] (15,19) [0.0]

(16,0) [0.0] (16,1) [0.0] (16,2) [0.0] (16,3) [0.0] (16,4) [0.0] (16,5) [0.0] (16,6) [0.0] (16,7) [0.0] (16,8) [0.0] (16,9) [0.0] (16,10) [0.0] (16,11) [0.0] (16,12) [0.0] (16,13) [0.0] (16,14) [0.0] (16,15) [0.01] (16,16) [0.83] (16,17) [0.0] (16,18) [0.0] (16,19) [0.0]

(17,0) [0.0] (17,1) [0.0] (17,2) [0.0] (17,3) [0.0] (17,4) [0.0] (17,5) [0.0] (17,6) [0.0] (17,7) [0.0] (17,8) [0.0] (17,9) [0.0] (17,10) [0.0] (17,11) [0.0] (17,12) [0.0] (17,13) [0.0] (17,14) [0.0] (17,15) [0.0] (17,16) [0.01] (17,17) [1.0] (17,18) [0.0] (17,19) [0.0]

(18,0) [0.0] (18,1) [0.0] (18,2) [0.0] (18,3) [0.0] (18,4) [0.0] (18,5) [0.0] (18,6) [0.0] (18,7) [0.0] (18,8) [0.0] (18,9) [0.02] (18,10) [0.06] (18,11) [0.02] (18,12) [0.0] (18,13) [0.0] (18,14) [0.0] (18,15) [0.0] (18,16) [0.0] (18,17) [0.0] (18,18) [0.31] (18,19) [0.03]

(19,0) [0.02] (19,1) [0.0] (19,2) [0.0] (19,3) [0.0] (19,4) [0.01] (19,5) [0.0] (19,6) [0.0] (19,7) [0.0] (19,8) [0.0] (19,9) [0.02] (19,10) [0.02] (19,11) [0.0] (19,12) [0.0] (19,13) [0.0] (19,14) [0.0] (19,15) [0.0] (19,16) [0.0] (19,17) [0.0] (19,18) [0.05] (19,19) [0.91]

};
\end{axis}
\end{tikzpicture}
    \centering
    \caption{\gls{fsel}, accuracy: 78.24\%}
    \label{fig:CM_FREL_vs_FSE_B}
    \end{subfigure}
     
    \setlength\abovecaptionskip{0.2cm}
    \caption{Confusion matrices for \FR and \gls{fsel} when trained in classroom, with monitor M1 and tested in office, with monitor M1. \vspace{-0.5cm}} 
    \label{fig:CM_FREL_vs_FSE}
\end{figure}

Figure \ref{fig:CM_FREL_vs_FSE} presents the confusion matrices of \FR and \gls{fsel} when trained with monitor M1 in the classroom and tested with monitor M1 in the office. It is evident from Figure \ref{fig:CM_FREL_vs_FSE_B} that the accuracy drop of \gls{fsel} is caused by a few activities which it finds difficult to distinguish. The top three activities which \gls{fsel} finds most difficult to distinguish are waiving while sitting (index 18), rotating (index 10) and eating (index 3). On the other hand, the top three distinct activities for \gls{fsel} are drinking (index 6), waiving (index 8) and phone call (index 13).

\begin{figure}[ht]
	\centering
	\includegraphics[width=.42\textwidth]{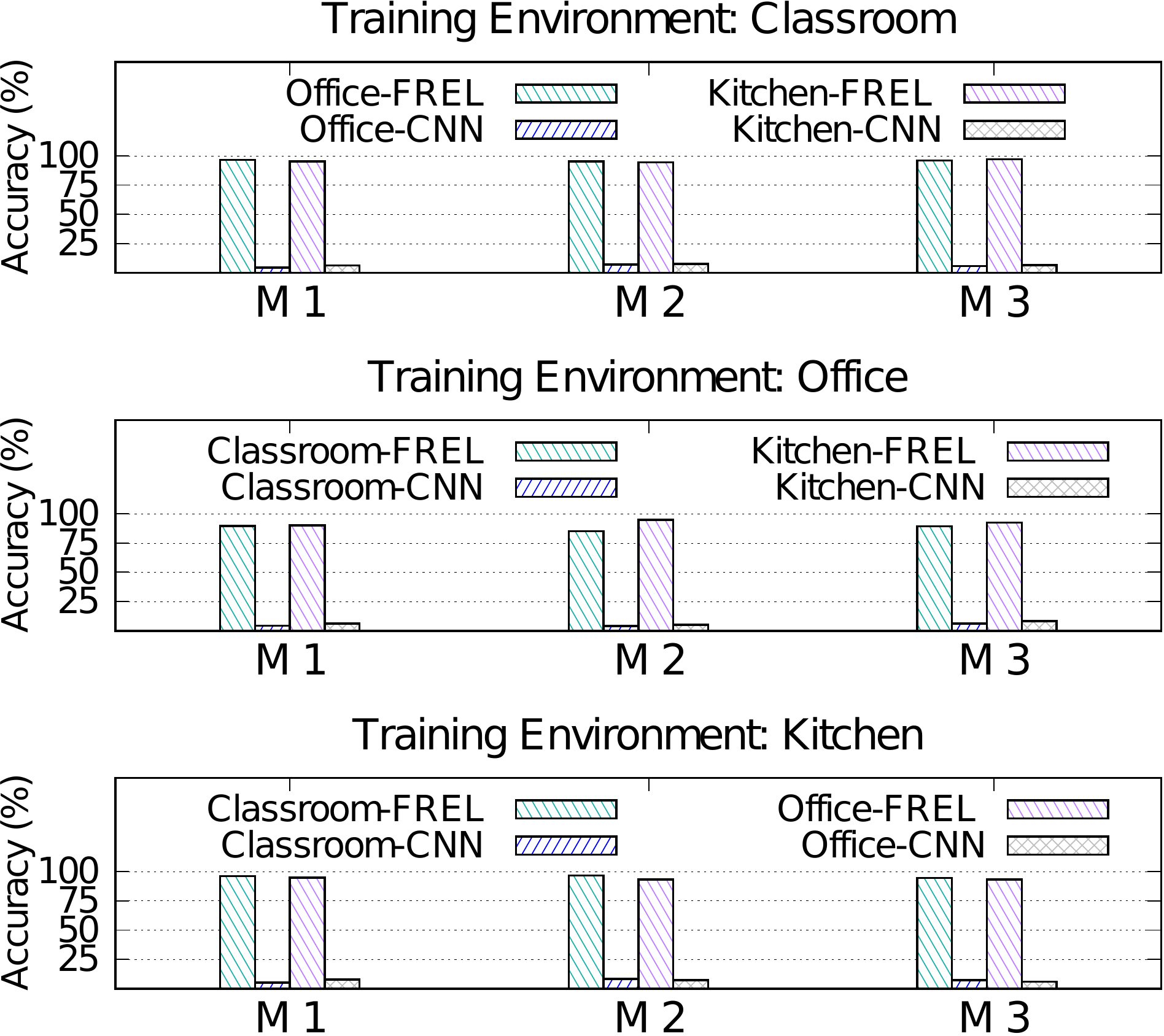}
	\caption{Performance of \FR as the subject identifier in untrained environments }
	\label{fig:cross_env_anomaly_cnn}
\end{figure}

\textbf{\FR in generalizing subject identification across environments:} \FR can also generalize the subject identification in new untrained environments, as presented in Figure \ref{fig:cross_env_anomaly_cnn}. In fact, it can achieve up to an accuracy of 96.53\% whereas the traditional \gls{cnn} only limits to 6.19\% compared to the 93.53\% for \FR on an average across the monitors and the environments.

\begin{figure}[ht]
	\centering
	\includegraphics[width=.42\textwidth]{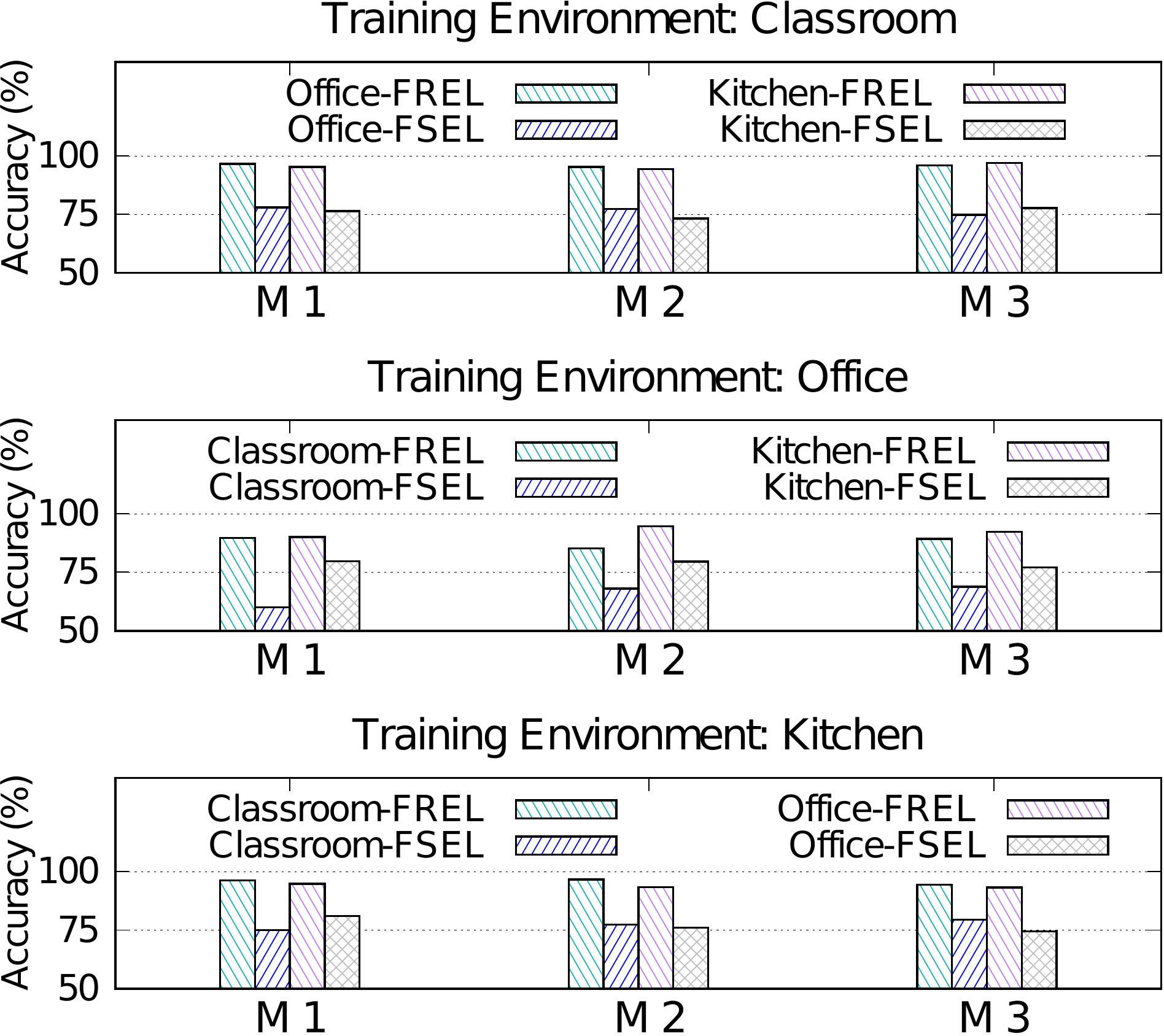}
	\caption{Performance comparison of \FR  and \gls{fsel} as the subject identifier in untrained environments }
	\label{fig:Free_vs_FSE_Anomaly}
\end{figure}

As depicted in  Figure \ref{fig:Free_vs_FSE_Anomaly}, when compared to \gls{fsel}, proposed \FR demonstrates an accuracy boost of  17.79\%, 17.93\%, and 17.55\% in classroom, office and kitchen, respectively. Thus, \FR signifies stable and reliable performances across the environments in generalizing the 'subject identification' phase of the learning also.  
\smallskip

\vspace{-0.5cm}

\begin{figure}[ht]
	\centering
	\includegraphics[width=.42\textwidth]{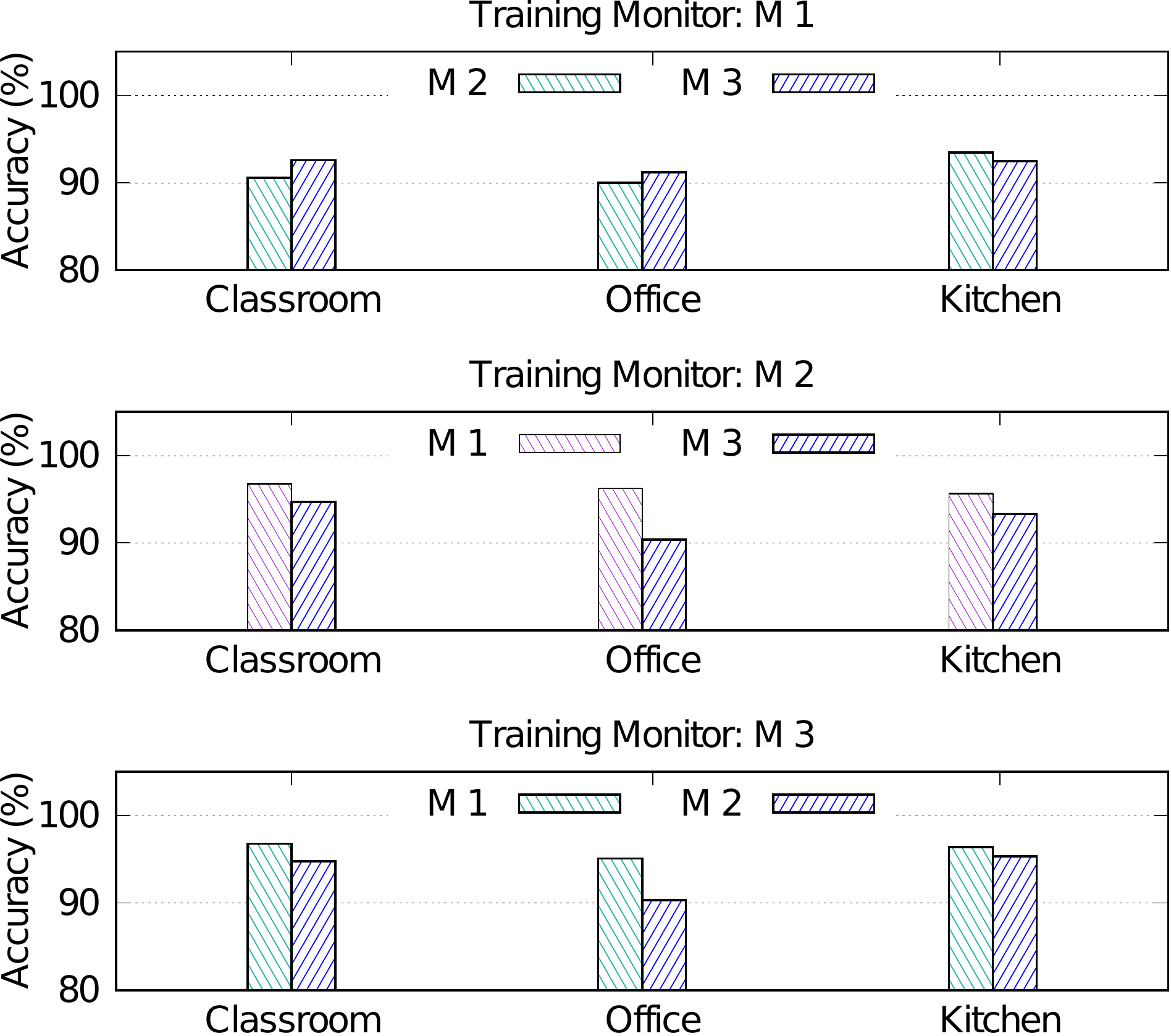}
	\caption{Performance of \FR with new untrained monitors.\vspace{-0.7cm} }
	\label{fig:cross_monitor}
\end{figure}

\vspace{0.5cm}
\textbf{\FR in generalizing the monitors across the environment:} It would be interesting  to see how a \FR model trained in one monitor can adapt and generalize to other monitors, thus generalizing new subjects of the same environment. It would further lessen the training time and drastically reduce the system deployment complexity. Figure \ref{fig:cross_monitor} presents the  performance of \FR with new untrained monitors across the three different environments. It can achieve an average accuracy of 91.71\%, 94.51\% and 94.78\% in generalizing the other monitors while trained with Monitor 1, Monitor 2 and Monitor 3, respectively. Thus \FR enables any system to be trained only on one monitor and deployed with n number of monitors with only 15s new data samples from each monitor.  

\section{Related Work}\label{sec:related_work}

A significant amount of research in Wi-Fi sensing has been performed over the last few years. The reader may refer to the following surveys for a good compendium of the state of the art \cite{ma2019wifi,tan2022commodity,hernandez2022wifi}. There have been several approaches to address the challenges of Wi-Fi sensing, which includes received signal strength indicator (RSSI) based approaches \cite{ssekidde2021augmented, dubey2021enhanced}, and passive Wi-Fi radar (PWR) \cite{yildirim2022multi, li2020passive}. Wi-Fi sensing leveraging beamforming feedback information (BFI) which can be captured from any Wi-Fi network is the state-of-the-art approach \cite{haque2023beamsense}. However, \gls{csi}-based Wi-Fi sensing is by far the most popular approach.

\gls{dl} has already been proven to be effective in various \gls{csi}-based Wi-Fi sensing applications, including human activity classification \cite{shalaby2022utilizing, yan2019wiact, ambalkar2021adversarial}, gesture recognition \cite{islam2020wi, gao2021towards}, health-monitoring \cite{pandey2022csi, kumar2022cnn, sharma2021passive}, human counting \cite{ibrahim2019crosscount, depatla2018crowd}, indoor localization \cite{abbas2019wideep}. To explore other \gls{dl}-based \gls{csi} sensing applications, we refer the readers to \cite{yang2022deep, ma2019wifi}. A few of the interesting works on human activity classification are briefly discussed below. Shalaby et al. \cite{shalaby2022utilizing} proposed four different \gls{dl} models, namely \gls{cnn} with Gated Recurrent Unit (GRU), gls{cnn} with GRU and attention, \gls{cnn} with a GRU and a second \gls{cnn}, and \gls{cnn} with Long Short-Term Memory (LSTM) and a second \gls{cnn}, achieved accuracy up to 99.46\%. However, these models only consider a single subject performing six activities and can not generalize to new untrained environments. MCBCAR by Wang et al. \cite{wang2021multimodal} used generative adversarial network (GAN) and semi-supervised learning to address the challenge of the non-uniformly distributed data due to environmental dynamics. Even though this work considers the dynamic change in the environment, the framework is not designed to adapt to new untrained environments and simultaneous multi-subject sensing. AFSL-HAR framework by Wang et al. \cite{wang2021robust} achieves a performance gain in recognizing the new activities with a few samples of new data through few-shot learning. 

Even though this work addressed the challenges of new scenario, activity or subject by fast adaptation with few new samples, the framework classify only one subject at a time in any environment. Ding et al. proposed WiLISensing \cite{ding2021device} to address the challenge of variations in activity locations in the same environment with few new data samples through Protonet. However, it is not evident how WiLISensing  would adapt to a completely new scenario or new subject. Ding et al. proposed RF-Net \cite{ding2020rf}, a meta-learning framework to adapt to new environments with few labeled data samples. However, the RF-Net's \gls{csi}-based sensing performance ranges in (70-80)\%. In contrast to the \gls{dl}-based systems, Abdelnasser et al. \cite{abdelnasser2018ubiquitous} designed WiGest for gesture classification based on mutually-independent distinguishable application actions which does not need any training. \textbf{To the best of our knowledge, we are the first to propose a framework for simultaneous multi-subject activity recognition with Wi-Fi sensing.}

\section{Concluding Remarks}
In this paper, we have proposed \FW, the first framework  for simultaneous multi-subject sensing based on Wi-Fi \gls{csi} sensing. In contrast to the existing approaches, \FW can classify the activities of multiple subjects in the same environment independently and simultaneously. An FSL-based algorithm: \FR is also proposed for fast adaptation to changing data distributions, making the system robust for the dynamic environment, which can generalize new environments, subjects with only a few new samples of data achieving an accuracy up to 98.94\%.  We have evaluated the efficacy of the overall design using extensive data collected in three different scenarios: classroom, office and kitchen, with three subjects performing 20 activities simultaneously. We demonstrate that \FW surpasses the traditional \gls{cnn}-based approach and a state-of-the-art \gls{fsl} model by 85\% and 20\% on average. 

\section{Acknowledgements}

This material is based upon work supported in part by the National Science Foundation (NSF) under Grant No. CNS-2134973 and CNS-2120447. The views and opinions expressed in this work are those of the authors and do not necessarily reflect those of the NSF.


\footnotesize
\bibliographystyle{IEEEtran}
\bibliography{mybib}
\end{document}